\documentclass[12pt,preprint]{aastex}

\newcommand{\simgt} {\,\hbox{\lower0.6ex\hbox{$\sim$}\llap{\raise0.6ex\hbox{$>$}}}\,}
\newcommand{\lya} {{\rm Ly}$\alpha$\ }
\newcommand{\msun} {{\rm\,M_\odot}}

\shorttitle{HII Bubble Growth during Reionization}
\shortauthors{Shin et al.}

\begin{document}

\title{COSMOLOGICAL HII BUBBLE GROWTH DURING REIONIZATION}
\author{MIN-SU SHIN, HY TRAC AND RENYUE CEN}
\affil{Princeton University Observatory, Peyton Hall, Princeton, NJ 08544-1001}

\begin{abstract}
We present general properties of ionized hydrogen (HII) bubbles and 
their growth based on a state-of-the-art
large-scale ($100$ Mpc/h) cosmological radiative transfer simulation.
The simulation resolves all halos with atomic cooling
at the relevant redshifts and simultaneously performs
radiative transfer and dynamical evolution of structure formation.
Our major conclusions include:
(1) for significant HII bubbles, the number distribution is peaked at
a volume of $\sim 0.6\ {\rm Mpc^{3}/h^{3}}$ at all redshifts.
But, at $z\le 10$, one large, connected network of bubbles dominates the entire HII volume.
(2) HII bubbles are highly non-spherical.
(3) The HII regions are highly biased with respect to the underlying matter distribution
with the bias decreasing with time.
(4) The non-gaussianity of the HII region is small when the universe 
becomes 50\% ionized. The non-gaussianity reaches its maximal near 
the end of the reionization epoch $z\sim 6$.
But at all redshifts of interest there is a significant non-gaussianity in the HII field.
(5) Population III galaxies may play a significant role in the reionization process. 
Small bubbles are initially largely produced by Pop III stars. 
At $z\ge 10$ even the largest HII bubbles have a balanced 
ionizing photon contribution from Pop II and Pop III stars,
while at $z\le 8$ Pop II stars start to dominate the overall ionizing photon production
for large bubbles, although Pop III stars continue to make a non-negligible contribution.
(6) The relationship between halo number density and
bubble size is complicated but a strong correlation
is found between halo number density and bubble size for large bubbles.
\end{abstract}


\section{INTRODUCTION}

Understanding the reionization process is a key challenge in cosmology.
Present observations provide extremely useful but still limited information with respect
to the epoch of reionization.
On the one hand, the absorption spectrum observations of high redshift 
quasars from the Sloan Digital Sky Survey (SDSS)
strongly suggest that the final reionization episode 
comes to completion at $z\sim 6$ \citep[e.g.,][]{becker01,fan02,barkana02b,
cen02}.
On the other hand, the Wilkinson Microwave Anisotropy Probe (WMAP) 
observations \citep{page06,spergel06} 
infer that the intergalactic medium 
is already significantly reionized by
some very early time, possibly $z\sim 15$.
In combination, it suggests that reionization process may be
quite complex and perhaps non-monotonic \citep[e.g.,][]{cen03,
haiman03,wyithe07a}.

In addition, some other observations and analyses have also provided useful
constraints on the reionization process.
\citet{theuns02,hui03} 
have shown that the observed temperature of the \lya forest
at redshift $z = 2-4$ requires that the cosmological reionization
occurred no earlier than $z = 9-10$,
although He II reionization somewhat 
complicates this constraint.
Analysis based on the SDSS quasar Stromgren sphere size
suggest that the neutral hydrogen fraction 
at $z\sim 6.3$ is a few tens of percent \citep{wyithe04,mesinger04}, 
whereas analyses based on the evolution of Lyman alpha emitters
from z = 6.5 to 5.7 imply that
a partially neutral IGM of neutral fraction of
$\sim 0.25$ is consistent with observations 
\citep{malhotra04,stern05,haiman05,malhotra06}. But the recent 
observations of Lyman alpha emitters begin to support the complete 
reionization at $z\sim6$ as the SDSS quasar observations imply 
\citep[e.g.,][]{kashikawa06,lidz07a}.

While a fully self-consistent reionization picture 
is far from being painted \citep[see][for a review]{barkana01,loeb06,fan06araa},
some key breakthroughs may rest in the redshift range $z = 7-15$,
where upcoming observations, including 21cm radio and CMB observations,
may be able to provide useful constraints.
To understand the reionization process several elegant analytical and semi-analytic 
models have been developed and predictions made 
\citep{miraldaescude00,barkana02a,furlanetto04,furlanetto05,furlanetto06,
zahn07,cohn07a}. 
These methods provide ways to economically explore the large parameter space 
and have provided very important insights with respect to 
HII bubble evolution as well as 
the evolution of global quantities such as the ionization fraction.
These methods, however, need to make necessary simplifying 
assumptions, which would limit the scope of their applicability or
the accuracy of the predictions.

We will take a complementary approach by making detailed numerical simulations
with fewer assumptions. In earlier simulations,
a choice is forced to be made between simulating a large volume
with very limited resolution or a sufficient resolution with too small volume
\citep[e.g.,][]{gnedin00,ciardi00,razoumov02,sokasian03,sokasian04,gnedin04}. 
However, both a large volume ($\sim 100$ Mpc/h) and an adequate resolution 
are necessary in order to follow the reionization sources
and sinks properly. A large simulation box is required because the sources in question
are extremely highly biased \citep{barkana04,trac06b}
and the bubbles have sizes of tens of Mpc prior to complete percolation 
\citep[e.g.,][]{furlanetto04}, 
whereas resolving the bulk of sources of halo mass $\sim 10^8\msun$ 
dictates, at least, the mass resolution of $\sim 10^6\msun$ and spatial resolution 
of a few kpc. With a unique hybrid (dark matter + baryons + stars + radiation) computational code
\citep{trac06b} 
we have crossed the threshold to be able to simultaneously simulate a large volume 
and have an adequate resolution to identify the bulk of the ionizing sources at 
high redshift as well as to have an approximate yet adequate treatment of radiation transfer.
Recent direct simulations performed by other groups \citep{iliev06a,zahn07}
have yielded very important results but the inability
to adequately resolve dark matter halos of $\sim 10^8\msun$ 
renders the results uncertain. Additional features of our method include 
(1) following simultaneously the evolution of structure formation and radiative transfer, 
instead of performing radiative transfer as a post-processing step \citep{iliev06b} 
or adding unresolved halos analytically \citep{mcquinn07}, 
(2) a self-consistent, albeit still uncertain, treatment of 
metal-enrichment process, hence the spatially varying transition from Population III 
(Pop III) to Population II (Pop II) initial mass function (IMF). Therefore, our 
simulations have more direct numerical treatments of radiation sources, small-scale clumping, 
and self-shielding than both post-processing and semi-analytic models.

In this paper, we analyze the evolution of HII bubbles in a $100$ Mpc/h simulation box. 
In particular, we study how HII bubbles grow and what physics determines the growth. Understanding 
the evolution of HII bubbles is closely related to predicting future observation results of the 
redshifted 21cm line and cosmic microwave background \citep[see][for a review]{fan06araa}. Among 
large volume reionization simulations the morphology of HII regions were studied in \citet{iliev06a,
zahn07,mcquinn07}. 
In \S2, the details of our simulation are described. 
Results of simulations are given in \S3, followed by discussion and conclusions in \S4. 
We use the following cosmological parameters based on the WMAP3 results \citep[see] 
[and references therein]{spergel06}: $\Omega_{m}$ = 0.26, $\Omega_{\Lambda}$ = 0.74, 
$\Omega_{b}$ = 0.044, h = 0.72, $\sigma_{8}$ = 0.77, and ${\rm n_{s}}$ = 0.95. 
Throughout the paper both length and volume are given in comoving units.

\section{SIMULATIONS}

Our simulation is generally based on the numerical methodology described in \citet{trac06b} and similarly utilized an N-body algorithm for dark matter, a star formation prescription, and a radiative transfer (RT) algorithm for ionizing radiation.  However, we have taken some simpler approaches in this initial step to satisfy the computational challenge of large volume, high resolution simulations of cosmic reionization.  In particular, we use an alternative RT algorithm and do not use a halo model for prescribing baryons and star formation.  Here we summarize the main components and describe the modifications.

In a ${\rm L = 100 ~ Mpc/h}$ simulation box, a high resolution N-body calculation for $2880^3$ dark matter particles on an 
effective grid with $11520^3$ cells is performed using a particle-multi-mesh code \citep{trac06a}.  The particle mass 
resolution is ${\rm 3.02 \times 10^{6}\msun}$ and approximately 33 particles make up a ${\rm 10^{8}\msun/h}$ 
halo.  The comoving grid spacing is 8.68 kpc/h and approximately 12 cells make up the virial volume of 
a ${\rm 10^{8}\msun/h}$ halo.  We identify dark matter halos in post-processing rather than during the 
course of the simulation.

We do not use the halo model of \citet{trac06b} for prescribing baryons and star formation.  Instead, an alternative 
approach is taken where we calculate the local matter density $\rho_m$ and velocity dispersion $\sigma_v$ for 
each particle.  The baryons are assumed to trace the dark matter distribution on all scales and we obtain 
the local baryon density $\rho_b=\rho_m(\Omega_b/\Omega_m)$ and gas temperature $T=\mu\sigma_v^2/(3k)$.  Star 
formation is only allowed to occur in particles with densities $\rho_m>100\rho_{\rm crit}(z)$ and temperatures $T>10^4$ K.  
This cut in the temperature-density phase-space effectively restricts star formation to regions within the virial 
radius of halos where efficient atomic line cooling allows the gas to dissipate energy and further collapse to very 
high densities.  We also differentiate between the first generation Pop III stars and the second generation Pop II 
stars by following the chemical enrichment of the interstellar medium and intergalactic medium as described in \citet{trac06b}.  

Our radiative transfer algorithm for ionizing radiation is based on a photon-advection scheme which is much less 
computationally expensive than ray tracing.  Particles, sources, and photons are collected on a RT grid with $360^3$ 
cells.  Each cell spans 278 comoving kpc/h and the RT time step is set by the light-crossing time.  For a source cell, 
the excess source photons are propagated to the 26 neighboring cells.  The advection is photon-conserving and the 
isotropic redistribution uses a weighting function that is proportional to $1/r^2$.  However, for a non-source cell 
with excess radiation, the advection is generally anisotropic.  For a HII region, photons originating from a central 
source propagate in the direction coinciding with decreasing radiation flux.  Therefore, we propagate photons by 
looking for gradients in the radiation field.  Consider a non-source cell with cell indices ${\rm (i, j, k)}$ and 
radiation density $n_\gamma{\rm (i, j, k)}$.  Radiation can propagate to an adjacent cell with 
indices ${\rm(i+di, j+dj, k+dk)}$ if either of the gradient terms,
\begin{eqnarray}
\Delta_+ & \equiv n_\gamma{\rm (i+di, j+dj, k+dk)}-n_\gamma{\rm (i, j, k)}, \\
\Delta_- & \equiv n_\gamma{\rm (i, j, k)}-n_\gamma{\rm (i-di, j-dj, k-dk)},
\end{eqnarray}
is negative.  The first gradient term indicates the downstream or expansion direction while the second 
term indicates the upstream or source direction.  For the 26 possible neighbor cells, we count how many 
cells satisfy the above criteria and redistribute the excess photons equally among them.  If none of the 
26 neighbor cells satisfy the above criteria, then the central cell is a convergent point and we redistribute 
the photons equally among the neighbors.  

For an isolated HII region with one central source, if one of gradient terms is false, the other is generally 
false too.  However, this may not be the case near the interface of merging or overlapping HII regions.  If 
a weaker radiation field is trying to expand into the vicinity of a stronger radiation field, then the first 
gradient term will be positive even though the second gradient term will be negative.  Therefore, it is 
necessary to use either of the gradient terms to determine the direction of radiation propagation.  In the appendix, we compare the photon-advection scheme with the ray-tracing scheme of \citet{trac06b} and show that this simpler approach correctly captures the radiative transfer for a significantly majority of the reionization epoch.

For each particle, we store 12 floating-point variables: three dimensional position, three dimensional velocity, 
matter density, baryon density, temperature, stellar fraction, HI fraction, HeI fraction, and HeII fraction.  We 
calculate the ionization and recombination for each particle individually rather than using the lower 
resolution density field defined on the RT grid.  This allows us to correctly account for the clumping factor 
and self-shielding of dense gas down to small scales of several comoving kpc/h.

The simulation was run at the National Center for Supercomputing Applications (NCSA) on a shared-memory SGI Altix with 
Itanium 2 processors.  We used 512 processors, 2 TB of memory, and approximately 80000 cpu hours.  With nearly 24 
billion particles, this is the largest cosmological N-body simulation run to date.  This is also the first 
reionization simulation with a ${\rm L = 100 ~ Mpc/h}$ simulation box that can 
directly resolve dark matter halos down to virial temperatures of $10^4$ K.

In post-processing, we have identified dark matter halos using a friends-of-friends (FoF) algorithm 
\citep{davis85} with a standard linking length of $b = {0.2 n^{-1/3}}$, where $n$ is the mean 
particle number density.  Figure 1 compares our mass functions with the analytic prediction of 
\citet{press74} and the empirical prediction of \citet{warren06}.  For redshifts 
$z\lesssim10$, our results are in very good agreement with recent works on the mass function of 
high redshift dark matter halos \citep{reed07, lukic07, cohn07b}.  For higher redshifts 
$z\gtrsim15$, we systematically under-resolve halos because our starting redshift of $z = 60$ is 
too late to capture the nonlinear gravitational collapse.  \citet{reed07} have suggested that 
simulations must start $\sim10-20$ expansion factors before the redshift at which results converge. 
We have found that we can accurately capture the formation of halos at $z\sim15-20$ with a smaller simulation 
starting at $z = 300$ \citep{trac06b}.

Our star formation prescription resembles those used in hydrodynamic simulations 
\citep[e.g.,][]{springel03} and we obtain very similar results.  In Figure 2, we compare 
the star formation rate $\dot{\rho}(z)$ with an analytical model calibrated using hydrodynamic 
simulations \citep{hernquist03}, corrected for our cosmology.  The overall shape is 
in very good agreement at all relevant redshifts.  At $z = 6$, our amplitude is 1.25 times larger 
and this difference is simply due to the fact that we have chosen a coarse value of $c_*=0.06$ 
for the star formation efficiency.  For our purposes, the overall amplitude of the star 
formation rate is unimportant since it is degenerate with the radiation escape fraction.  We 
have correspondingly chosen a radiation escape fraction of $f_{\rm esc}=0.15$ in order to have 
the redshift of complete overlap occur at $z\approx6$.

\section{RESULTS}

The first reionization region appears when regions of dense baryons turn on star formation. 
Once the radiation sources begin ionization around them, the global ionization regions appear 
as shown in Figure 3. At the early stage of reionization, isolated ionization regions are 
easily found (see z = 11.228 in Figure 3).
As more radiation sources develop, the isolated bubbles get  
connected together along filaments.
The process of HII bubble merging is 
very complex and difficult to treat without detailed simulation.
Visually, in Figure 3, the computed  HII bubbles 
do not appear to be close to spheres,
as will be confirmed quantitatively below.
After $z \sim 10$
one large connected network of HII bubbles begins to dominate the simulation volume,
the HII percolation process
enters the ``cannibalistic" phase where 
the dominant HII bubble rapidly swallows
other HII bubbles, as evident 
in the z = 9.3, and 7.2 panels. 
In the following subsections, we will present quantitative results.

\subsection{Size of HII bubbles}

\subsubsection{Characteristic size}

The size of HII bubbles is a basic quantity that has
received significant attention
\citep{furlanetto05,furlanetto06,iliev06a,mcquinn07,zahn07}. 
We define HII bubbles as follows.
First, we mark all the cells 
that are ionized at $>50$\% level.
Then, a bubble is defined by all such cells that are connected 
at least by one face;
this is practically done by 
grouping cells using  a friends-of-friends (FoF) algorithm
with a linking length equal to simulation data ouput cell size, i.e. $\sim 0.14\ {\rm Mpc/h}$. 
In addition, at least two cells are needed to make up a single bubble.

Since the shape of HII bubbles is not close to a sphere as shown in \S3.2, 
we present, in Figure 4, the volume 
distribution of the found HII bubbles instead of the radius of a sphere.
Note that in previous 
analytical studies HII bubbles are often assumed to be 
spherical \citep[e.g.,][]{pritchard06} or have a characteristic size
\citep[e.g.,][]{mcquinn07}. 
We present the size distribution of z = 5.7, 6.3, 8.9, 13.5, 17.2, and 20.6.

In this paper, we define the 
characteristic size of bubbles based on their number fraction not volume fraction as used 
in \citet{iliev06a}. Therefore, our characteristic sizes well describe the existence of many small 
non-connected bubbles. Meanwhile, 
the definition of \citet{zahn07} is affected by volume occupied by HII bubbles because it measures 
an ionization fraction within a certain smoothing radius. 

The volume distribution by the number fraction shows that there are three characteristic sizes of the HII 
bubbles before the simulation box is dominated by a single bubble.
 The characteristic volumes are 0.6,\ 0.03,\ and 0.006 
${\rm Mpc^{3}/h^{3}}$. These characteristic volumes do not change even though more regions become 
ionized. But the least volume of the possible bubbles is $\sim 0.005\ {\rm Mpc^{3}/h^{3}}$ because 
our bubble finding method defines the smallest bubble as at least two connected simulation cells. 
Therefore, the smallest-volume peak in the HII bubble distribution
may be limited by our radiation cell size and should be considered as an upper limit 
for the peak at the smallest scales. 

Other two characteristic sizes of $\sim 0.03$ and $\sim {\rm 0.6\ Mpc^{3}/h^{3}}$ are 
real features of the size distribution.  The peak at $\sim 0.03\ {\rm Mpc^{3}/h^{3}}$ represents 
HII bubbles that are not limited by our radiation transfer resolution and should be real.
The peak at $\sim 0.6\ {\rm Mpc^{3}/h^{3}}$ 
probably represents typical mature HII bubbles produced by individual galaxies
(probably with their satellite galaxies),
which may subsequently merge with other bubbles.
This peak is maintained at all times until
when the percolation permeates the entire volume to complete
the reionization, as seen from the difference
between z = 6.3 and z = 5.7 in Figure 4.
The properties of these characteristic sizes 
will be further analyzed and understood in \S3.1.3 and \S3.1.4.

The appearance of few dominant large bubbles is easily found by the change of volume fraction for 
different sizes of bubbles. As shown in Figure 4, the volume fraction is dominated by 
few large bubbles after z = 13.5. This result is also visually recognized in Figure 3. 
In the plot of z $\sim$ 11.2, large bubbles begin to appear while many small bubbles 
around newly formed sources dominate the number fraction in Figure 4. Therefore, the 
characteristic sizes of 0.6,\ 0.03,\ and 0.006 ${\rm Mpc^{3}/h^{3}}$ can be maintained 
until the scale of large bubbles is big enough to percolate small bubbles that are distributed 
between large bubbles.

The volume distribution of HII bubbles in \citet{iliev06a} can be compared to our results. 
Both simulations use the same definition of HII bubbles and the bubbles 
are found by FoF method. The first difference is that in our simulation 
large bubbles of $\simgt 10^{4}\ {\rm Mpc^{3}/h^{3}}$ appear later than z $\sim$ 15 when 
the global volume-weighted ionization fraction is about 0.04. But 
\citet{iliev06a} found the existence of those large bubbles around z $\sim$ 15 when 
in their simulation the global volume-weighted ionization fraction is 0.05. This 
difference is also related to different characteristic sizes that found in our simulation. 
In our simulation, most HII bubbles are much smaller than $\simgt 10^{4}\ {\rm Mpc^{3}/h^{3}}$. 
Even the found characteristic size of intermediate-size bubbles is $\sim 0.6\ {\rm Mpc^{3}/h^{3}}$. 
\citet{iliev06a} generally shows larger bubbles for all ranges of the global ionization fraction 
and the formation of large bubbles at 
the earlier times than ours. Moreover, the largest fraction of bubbles is explained by 
intermediate-size bubbles in \citet{iliev06a} while our results show the dominance of small bubbles 
for almost all simulation time. 

We think this is probably due to the difference in the adopted 
cosmological model in which the model used by \citet{iliev06a} has a higher $\sigma_8$
and the fact that they require the universe to be ionized much earlier such that
their computed $\tau_e$ matches the results from the first-year WMAP results. The low mass resolution 
of \citet{iliev06a} also does not permit the formation of small bubbles because it cannot 
capture the formation of low-mass DM halos \citep{zahn07}. Finally, the implementation of 
baryon physics like ionizing sources in their simulation is quite different from our prescription 
given in \S2. The importance of this difference is explained in \S3.1.3 and \S3.1.4.


In Figure 4, we also present the bubble size distribution based on the definition of \citet{zahn07} 
for comparison. 
The same bubble boundary threshold of 90\% is adopted in this measurement. For similar 
global volume-weighted ionization fractions as \citet{zahn07}, the size distribution from our 
simulation has its peak at the smaller radius than their size distribution. The existence 
of many resolved small bubbles results in the difference of the size distribution even though 
we adopt the same definition as \citet{zahn07}. For example, we find two peaks of size distribution 
at $\sim 0.1$ and $0.5$ Mpc/h for the volume-weighted ionization fraction of 0.12, while 
\citet{zahn07} found a single peak at $\sim 0.7$ Mpc/h for the volume-weighted ionization 
fraction of 0.11. Comparing the results for the volume-weighted ionization fraction 
of 0.52, our distribution shows a single peak size, but the size is $\sim 2.5$ Mpc/h that is 
smaller than $\sim 4.5$ Mpc/h of \citet{zahn07}. In particular, the resolved small bubbles makes the 
peak probability of the size distribution, i.e. the height of the peak, insensitive to the change 
of a global ionization fraction. Therefore, even with the same definition of the bubble size, 
our result shows a slightly different result although a general result is consistent with each other.

\subsubsection{Bubble merger history}

As it turns out, the percolation of bubbles is also related to the transition of major 
ionizing photon production from Pop III to Pop II stars.

To shed light on the overall reionization process,
we follow the change of ionizing environment at four different location 
in the simulation box that are occupied by four individual bubbles that 
all started at $z\sim 21$, shown in Figure 5. In the plot, we track the 
size change of bubbles that occupy the four locations. 
We see that these bubbles initially, formed at about the same time but 
at four different locations, display a variety of histories, clearly due to the rich, disparate 
structure formation histories and the evolution of ionizing environment. 
One striking and very clear signature is some intermittent, rapid
downturns in the HII volume during an individual location's evolution history.
This is due to a rapid transition 
of the IMF from Pop III to Pop II in the bubble, which in general occur at 
the redshift range of $z = 10 - 15$.
This is consistent with the overall contributions of Pop II stars and and Pop III stars
as a function of redshift shown in  Figure 6,
where we see that ionizing photons from Pop II stars begin to become more dominant between z = 10 and 15. 
But this transition occurs spatially at different time such that each bubble position has 
a slight different history. 
We note that this transition time is not the same as the epoch when star formation rate is dominated by 
Pop II population, since Pop III stars are more efficient producers of ionizing photons than Pop II stars
for a given amount of star formation rate. \citet{mcquinn07} pointed out that the nature of the ionizing sources 
plays an important role in shaping the bubbles and our results
quantitatively confirm their conclusion by being able to perform 
a detailed calculation that allows for a spatially resolved transition 
from Pop III to Pop II stars.

Merger history of the bubbles is also consistent with the changing total number of bubbles and 
the dominance of the largest bubble over small bubbles after z $<$ 10. As percolation of the 
bubbles proceeds, the total volume of HII bubbles increases while the number of individual bubbles 
decreases as shown in Figure 7. That is, percolation of the bubbles becomes so important that 
the largest network of connected bubbles dominates the simulation box. Although the number of 
small bubbles also begins to decrease after z $<$ 10, the number fraction of small bubbles increases 
because the size of the small bubbles is not large enough to be connected to other bubbles (see 
Figure 4).

\subsubsection{Dark matter halo and bubble size distribution}

Basically, the clustering of radiation sources affects the percolation 
domination and the spatially varying different transition epoch from Pop III star formation 
to Pop II star formation. Because the two small characteristic size peaks found in Figure 4 
are maintained by the formation of new radiation sources, the correlation between the bubble sizes 
and the DM halos can be expected to explain a physical reason behind the bubble size distribution.

In Figure 8 we show the number density of halos as a function of HII bubble volume. 
All DM halos whose centers are inside the bubble are said to belong to the bubble. 
We separate the member DM halos to three separate mass ranges, ${\rm M_{halo} < 
10^{8}\ M_{\odot}/h,\ 10^{8}\ M_{\odot}/h \leq M_{halo} < 10^{9}\ M_{\odot}/h}$, and 
${\rm 10^{9}\ M_{\odot}/h \leq M_{halo}}$.
We see that there is an upturn of halo number density for all mass ranges for the smallest bubbles.
This indicates that these bubbles are rapidly expanding due to newly formed sources while the 
number of halos stays approximately constant.  These smallest bubbles do not contain 
DM halos of ${\rm M_{halo} > 10^{9} M_{\odot}}$, corresponding to the peaks of 
${\rm \sim 0.11\ and\ 0.19\ Mpc/h}$ 
as found in \S3.1.1.
The halo number density decreases, as one moves to the right,
and then levels out at $\sim -1$ to $0$ of the displayed x-axis.
The most interesting feature is that there is a sudden rise 
of halo number density of all masses toward the largest bubbles,
and points to the direction that percolation is expected to continue 
with time, consistent with our analysis of the actual bubble size evolution. 
The boundary between the percolation dominance and the source formation dominance is 
the characteristic volume of ${\rm \sim 0.6\ Mpc^{3}/h^{3}}$. 
This feature is described as the beginning of the flat 
DM halo density distribution in Figure 8. A large drop in the minimum number density is 
caused by difference of bubble age among DM halos of the same mass. For example, 
the same mass DM halos of ${\rm 10^{8.5}\ M_{\odot}/h}$ collapse at different epoch and 
it results in the differently matured HII bubbles around them. So, the number density 
of DM halos can have a large dispersion, in particular, for the intermediate-mass DM halos.

As characteristic sizes of HII bubbles do not change much until z $\sim$ 6, 
the DM halo distribution does not change as shown for z $\sim$ 8 and 10. So, we find that the size distribution 
of HII bubbles are mainly regulated by an overall mass density that determines the evolution of baryons.
The latter property indicates highly biased galaxy formation and
suggests that stellar mass density is expected to rapidly rise in high density regions.
Whether this property may be preserved requires more detailed analysis.

\subsubsection{Dependence of bubble size on ionizing photon production}

The rate of ionizing photon production per unit volume is likely
the main physical factor to determine the evolution of HII bubbles,
while the broad correlation between star formation rate and DM halo
density has likely produced some of the interesting 
features presented in the previous sub-section,
because the latter is essentially proportional to 
the total, integrated ionizing photon number density.
An ionizing photon production rate inside HII bubbles is estimated 
to be proportional to star formation rate times 
the ionizing photon production efficiency that
is a function of the IMF.
Figure 9 shows the ionizing photon production rate, separately from
Pop III stars and Pop II stars, as a function of the bubble size 
at $z=8$ and $z=10$, respectively. The flat part of the average 
ionizing photon production rate represents the value which is 
approximately set by the average baryon number density and 
recombination rate inside the bubbles. 
It is clear that for the smallest bubbles with size $\le 0.3$ Mpc/h 
Pop III stars dominate the ionizing photon production rate at both $z=10$ and $z=8$,
indicating that these bubbles are produced by 
the first generation of stars formed in those locations
which are likely relatively removed from large halos where
star formation has gone through more than one generation and
the gas has been significantly enriched with metals,
consistent with Figure 8.
We expect that Pop III galaxies may be present in small bubbles
until $z\sim 6$.
For the large bubbles (size $\ge 10$ Mpc/h)  
the contributions from Pop III stars and Pop II stars 
seem balanced at $z\sim 10$.
But, by $z\sim 8$, these large bubbles begin to be dominated by
Pop II stars, although Pop III stars continue to make
significant contributions of ionizing photons.

We show the careful 
consideration of this transition is needed to understand the reason why we find the bubble size 
distribution given in Figure 4. 
We believe that the overall evolution of HII bubbles 
depends on detailed modeling 
of Pop II and Pop III stars.
A neglect or non-detailed treatment 
\citep[e.g.,][]{zahn07,furlanetto05,iliev06b} 
of the Pop III stars 
would translate to a significant
change in the characteristics of bubble evolution.

\subsection{Shape of HII bubbles}

The shape of HII bubbles are far from a sphere unlike what is assumed in most analytical studies. Our 
result is the first effort to present a quantitative measurement of HII bubble shape in the research of 
cosmic reionization. 
In order to quantify the shape of the bubbles as ellipsoid, first, we calculate 
the inertia tensor: 
\begin{equation}
I_{ij} = \Sigma (x_{i} - x_{i}^{c}) (x_{j} - x_{j}^{c}),
\end{equation}
where $x$ is the coordinate of every cell that is included in each HII bubble, and $x^{c}$ is 
a geometrical center of the bubble. From the above defined $I_{ij}$, we obtain the square roots of 
the three eigenvalues, $a > b > c$. We measure the shape of the bubbles that have at least six member 
cells. Therefore, the smallest volume of the bubble is $\sim 0.016 {\rm \ Mpc^{3}/h^{3}}$. 
The shape is expressed as two parameters $e_{1}$ and $e_{2}$:
\begin{equation}
e_{1} = \sqrt{1 - \frac{b^{2}}{a^{2}}},\ e_{2} = \sqrt{1 - \frac{c^{2}}{b^{2}}}.
\end{equation}
After percolation becomes a main process in growth of HII bubbles, the above measurement of 
shape is useless. At late stage of reionization, the complicatedly connected bubbles are not 
similar to ellipsoids. Moreover, neutral hydrogen regions can be surrounded by HII bubbles 
such as a grape-like structure. Therefore, we derive the $e_{1}$ and $e_{2}$ for bubbles 
before the largest bubble is comparable to the simulation volume.

The bubble volume-weighted shape distribution is estimated in the following way:
\begin{equation}
P(e_{1}, e_{2}) = \int_{V} P(e_{1}, e_{2} | V) P(V) dV.
\end{equation}
Figure 10 shows 
the concentration of volume-weighted HII bubbles in $(e_1,e_2)$ plane.
Note that a spherical bubble would have $e_1=e_2=0$.
In Figure 11 we show the bubble shape as a function of 
bubble size, with the largest bubble separately shown as the star symbol,
for $z=13.8$ and $z=10$.
The shape of the bubbles follows a complicated dependence on their sizes. 
The shapes of small bubbles are, however, less accurately calculated because of limited resolution.
Clearly, most of the volume in HII bubbles 
is far from spherical.
The fact that $e_1$ is quite close to $1$ and $e_2$ has a wide range indicates
that bubbles tend to have shapes ranging from a cigar to a ruler.
Note that at $z<12$ the shape distribution is basically dominated by the largest bubble.
For example, at z = 10.0, the most probable shape parameter is close to $e_{1} \sim  
0.9$ and $e_{2} \sim 0.6$ of the largest bubble, 
corresponding to axial ratios $b/a=0.4$ and $c/b=0.8$.
This high complex non-sphericality would introduce 
substantial inaccuracies in calculations based on
analytic modeling of observables of the high-redshift universe \citep[see][for a review]{fan06araa}. 
The complicated shape of HII bubbles 
is largely determined by the clustering of ionizing sources, 
as bubbles grow around large-scale structure, primarily filamentary networks.
Figure 12 gives the three orthogonal projections of a single randomly chosen bubble 
which show characteristic of bubbles resulting from mergers of smaller bubbles.

\subsection{Non-Gaussianity of reionization}

We sample 5000 spheres randomly placed 
in the simulation box and measure the mass-weighted ionization fraction of the spheres. 
That is, we measure $x_{bubble}$ = (mass of HII) / (mass of H) for each sampled sphere. It is a 
slightly different way to find a PDF compared to that of \citet{iliev06a}
in which they sample independent sub-cubes in their simulation box.
From these sampled spheres, 
we estimate mean ($\mu$), variance ($\sigma^{2}$), skewness ($\gamma_{1}$), and kurtosis 
($\gamma_{2}$). 
We use three different sphere sizes of comoving radius $5$, $10$, and $20\ {\rm Mpc/h}$, respectively.
Figure 13 shows measurements 
of the underlying HII probability distribution functions.
First of all, we see that the mean ionization fraction reaches
50\% at $z\ \sim\ 9.5$ (hereafter, ${\rm z_{50\%}}$).
Because the size of sampled spheres is much larger than the characteristic bubble sizes, 
the measured $\mu$ reflects the global mean ionization fraction. 
The largest dispersion of ionization fraction appears at 
$z\ \sim\ 8.9$,
close to ${\rm z_{50\%}}$.
It is easily seen that 
the degree of non-Gaussianity decreases with increasing size of the sphere,
as shown in all three bottom panels,
as expected.
While the variance starts to fall at $z< 8.9$,
the non-Gaussianity as measured by
$\gamma_{1}$ and $\gamma_{2}$ 
continue to rise, peaking at $z\sim 5.7$ when the largest bubble starts to fill
the entire simulation box.
Interestingly, it might be of some comfort for analytic modelers
that at ${\rm z_{50\%}}$, the HII field 
is weakly non-Gaussian with $\gamma_1=(0.22, 0.17, 0.01)$ and $\gamma_2=(-0.66, -0.26, -0.46)$,
for $r=(5,10,20)$ Mpc/h, respectively. But $\gamma_1$ and $\gamma_2$ are most close to 
0 at $z \sim 9$ and $\sim 10$, respectively. Because the boundary between fully neutral and fully 
ionized regions is much more smaller than the sizes of the sampling spheres, the measured 
non-Gaussianity is quite small around ${\rm z_{50\%}}$ \citep{wyithe07b,lidz07b}.

\subsection{Power spectrum of ionized hydrogen density fluctuation}

We present the 3D power spectrum of ionized hydrogen density fluctuation field 
(${\rm \delta_{HII}}$), neutral hydrogen density fluctuation field (${\rm \delta_{HII}}$), 
and matter density fluctuation field (${\rm \delta_{mat}}$) in Figure 14. We note that 
the power spectrum is not measured for ionization fraction fluctuation but for ionized density 
fluctuation. The dimensionless power spectrum is defined here as

\begin{equation}
\Delta_{k}^{2} = {\rm \frac{V}{(2 \pi^{2})} 4 \pi k^{3} ~ P(k)}.
\end{equation}

We see that the ionized hydrogen density field shows a strong bias with respect to 
the matter power spectrum,
where the bias is the strongest on small scales 
and decreases toward large scales.
In general, the bias of ionized regions always decreases with time,
whereas the bias of the neutral regions does the opposite.
We find that ${\rm \Delta_{HI}^2}$ matches to that of ${\rm \delta_{mat}}$ at the early time when
most hydrogen is still neutral, as expected,
whereas 
${\rm \Delta_{HII}^2}$ becomes equal to that of ${\rm \delta_{mat}}$ when
the universe gets completely ionized.
This is in agreement with the underlying physics that the earlier bubbles are produced
by highly biased galaxies formed inside DM halos of high $\sigma$ peaks,
whereas only those highly biased hence large galaxies host
most of the neutral hydrogen when the universe becomes highly ionized.
It is interesting to note that
at around ${\rm z_{50\%}}$ 
${\rm \Delta_{HI}^2}$ and ${\rm \Delta_{HII}^2}$ 
appear to have roughly the same shape,
although the latter has somewhat more power than the former,
while both have more power than the total matter,
suggesting a significant non-gaussian distribution of both.
We also note that there are significant difference
between the nonlinear (actual) power spectrum of total matter and
the linear power spectrum at $k\ge 0.3$ at $z\le 10$, indicating 
the necessity to take nonlinear effects into account.

\section{CONCLUSIONS AND DISCUSSION}

Using the state-of-the-art, largest cosmological simulation of box size $100$ Mpc/h 
with detailed radiative transfer of ionizing photons,
we compute the evolution 
of HII bubbles during cosmological reionization from $z\sim25$ to $z\sim6$.
Our simulation resolves galaxy sources that produce most of the 
ionizing photons in the universe in the concerned redshift range.
Here are a few major findings from our analysis.

(1) We find that, for significant HII bubbles,
their number distribution is peaked at
a volume of $\sim 0.6$ ${\rm Mpc^{3}/h^{3}}$ at all redshifts.
But, at $z\le 10$, one large, connected network of bubbles dominates the entire HII volume.

(2) HII bubbles are highly non-spherical. This result is not totally unexpected, 
since one is quite familiar with the generic filamentary nature 
of cosmic structure formation.
The mergers of HII bubbles blown by galaxies formed along filaments
consequently produce filamentary (fatter) HII bubbles.

(3) The HII regions are highly biased with respect to the underlying matter distribution
with the bias decreasing with time as more and more less biased structures
begin to dominate the ionizing photon production rate. The opposite is true for
the neutral hydrogen region in the sense that 
the bias of the neutral regions increases with time.

(4) The universe becomes 50\% ionized at redshfit $z\sim 9.5$,
when the HII region is actually the least non-gaussian 
in the redshift range concerned.
The non-gaussianity of the HII region reaches its maximal near the end of the reionization epoch $z\sim 6$.
But at all redshifts of interest there is a significant non-gaussianity in the HII field.

(5) Population III galaxies play a very important role in the reionization process.
With our spatially resolved treatment of metal enrichment,
we show that small bubbles are initially largely produced by Pop III stars.
At $z\ge 10$ even the largest HII bubbles have a balanced 
ionizing photon contribution from Pop II and Pop III stars.
At $z\le 8$, however, Pop II stars start to dominate the overall ionizing photon
budget for large bubbles, although Pop III stars continue to make
a non-negligible contribution.

(6) The relationship between halo number density and
bubble size is complicated, although there is a strong correlation
for large bubbles in which larger bubbles tend to have higher halo number density.
In addition, we find that only the large bubbles (size $\ge 1$ Mpc/h) 
contain halos of mass in excess of $10^9\msun$/h.

Some of the results depend on modeling the formation of Pop III stars and the transition 
from Pop III to Pop II star formation that needs more careful studies 
\citep[e.g.][]{schneider06,smith07}. First 
of all, the correct ionizing photon generation rate from the unit mass of Pop III stars or Pop II  
stars is as important as a careful treatment of recombination process in reionization simulation. 
As shown in \S3.1.4, the formation of small bubbles is closely connected to the ionizing efficiency 
of the Pop III stars that is subject to the uncertain IMF of the Pop III stars.
Our reionization simulation is the first trial to simulate the transition of Pop III star formation to 
Pop II star formation in a large simulation box of 100 Mpc/h. Improving our understanding 
of chemical enrichment during reionization \citep[e.g.][]{greif07}, the transition of 
a dominant radiation source will be more certainly simulated in future.

In addition to the uncertain physics of Pop III population, the numerical treatment of 
self-shielding and recombination has to be improved in future simulations. We used 
information of simulation particles for calculating the recombination rate that has 
better resolution than grid-based approach. But out approach also has limitation of 
resolving small-scale physics like self-shielding. That must be improved in the reionization 
simulation with a sub-grid model.

\acknowledgments
We thank Paul Bode for allowing us to use his FoF code and Marcelo Alvarez for 
his bubble size measurement code. We also like to thank the referee for 
many constructive comments that improved this paper. This research is 
supported in part by grants AST-0407176 and NNG06GI09G. 
HT is additionally supported in part by NASA grant LTSA-03-000-0090.

\appendix
\section{RADIATIVE TRANSFER:  PHOTON-ADVECTION VERSUS RAY-TRACING}

We compare the photon-advection scheme with the ray-tracing scheme of \citet{trac06b} by 
applying these RT algorithms to two reionization simulations with identical initial 
conditions and source prescriptions.  The test is conducted in a 25 Mpc/h box with 
$720^3$ dark matter particles and $90^3$ RT cells and has the same spatial, mass, 
and temporal resolution as the large 100 Mpc/h simulation.  

We find that the photon-advection scheme produces very similar results in terms of 
the spatial and temporal evolution of HI for a substantial majority of the 
reionization epoch.  Figure 15 shows very good agreement in the redshift evolution of 
the volume-averaged HI fraction $f_{\rm HI}$ and deviations are only found when the box 
is already significantly ionized.  The timing starts to differ at late stages near 
complete overlap.  There are small delays of $\Delta z\sim0.1$ and 0.2 when the neutral 
fraction drops to 0.1 and 0.01, respectively.

In order to quantify the agreement in the spatial evolution of HI, we 
cross-correlated the $f_{\rm HI}(x)$ fields from the two RT simulations and 
plotted the results in Figure 16.  From the two power spectra, $P_{\rm adv}(k)$ 
and $P_{\rm ray}(k)$, and the cross-power spectrum $P_{\rm adv-ray}(k)$, we 
can define the bias function,
\begin{equation}
b(k)\equiv\sqrt{\frac{P_{\rm adv}(k)}{P_{\rm ray}(k)}},
\end{equation}
and cross-correlation function,
\begin{equation}
r(k)\equiv\frac{P_{\rm adv-ray}(k)}{\sqrt{P_{\rm adv}(k)P_{\rm ray}(k)}}.
\end{equation}
and use them to quantify the statistical correlation between the two fields. 
In both simulations, the neutral fraction drops to 0.5 and 0.25 at $z=8.2$ and 7.5, 
respectively.  At these two stages of reionization, the bias $b(k)$ is very close to 
unity and only deviates by a maximum of $\sim0.1$ near the RT grid Nyquist frequency 
$k=11.3$ h/Mpc. Similarly, the cross-correlation $r(k)$ shows very good agreement 
even down to the smallest scale. In the ray-tracing simulation, the neutral fraction 
drops to 0.1 at $z=7.1$, but $f_{\rm HI}$ is slightly higher by $0.04$ in the 
photon-advection case.  At this redshift, the bias and cross-correlation 
functions show that the two fields differ appreciably at all scales. 
However, when the two simulations are compared at the same neutral fraction 
of 0.1, the correlation is much better, particularly on large scales. 
The deviations from unity at the smallest scales are due to the appearance of 
new sources in the photon-advection simulation taken at a slightly later redshift.

In summary, we find that the photon-advection scheme correctly captures
the reionization process up until it is $\sim75\%$ completed by volume. At
earlier stages, the radiation field is highly nonuniform, even within the HII
regions, and the propagation of photons in the direction of decreasing radiation
flux is a good description. However, at later stages of reionization near
complete overlap, the radiation field is much more uniform and the weak
gradients in the radiation density do not provide accurate directions for photon
propagation. We conclude that the photon-advection scheme provides a cost-effective
approach to radiative transfer for a significant majority of the reionization epoch.
However, the stages just before and after reionization should be simulated
using more accurate approaches like ray-tracing.  Our results in this
paper are valid at all stages when the the ionization fraction, rather
than the redshift, is used as an indicator of the progress of reionization.

\begin{figure}
\plotone{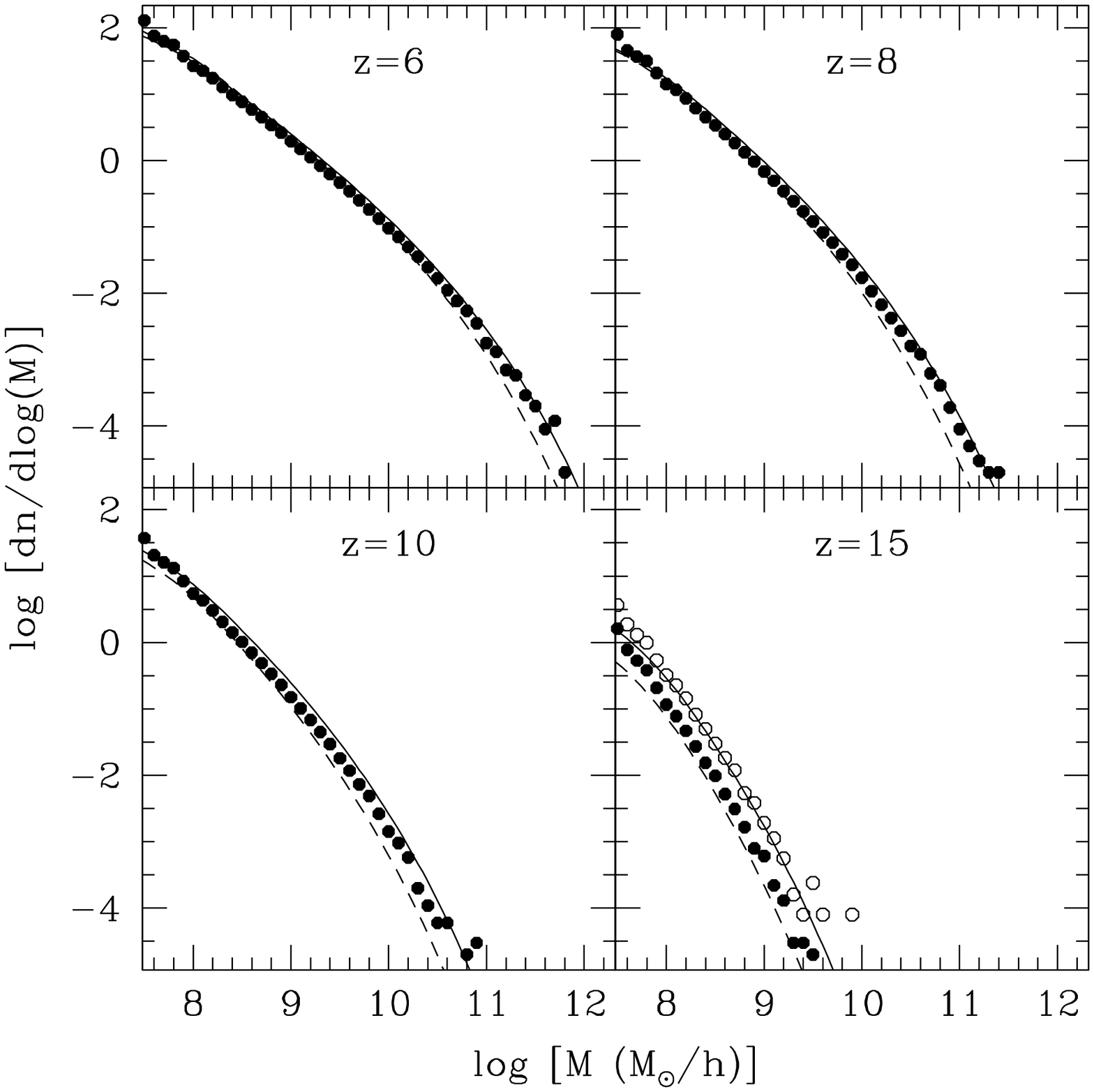}
\caption{Dark matter halo mass functions. Dark matter halos are identified 
using a friends-of-friends algorithm with a standard linking 
length of 0.2 times the mean inter-particle spacing. Our mass functions are in better agreement with 
\citet{warren06} {\it (solid)} than with \citet{press74} {\it (dash)} for $z\lesssim10$. 
At $z\gtrsim15$, we under-resolve the halos because of the late starting redshift of $z = 60$. 
A smaller simulation starting at $z = 300$ correctly captures the abundance of high redshift 
halos {\it (open circles)}.}
\label{fig:halo_mass_func}
\end{figure}

\begin{figure}
\plotone{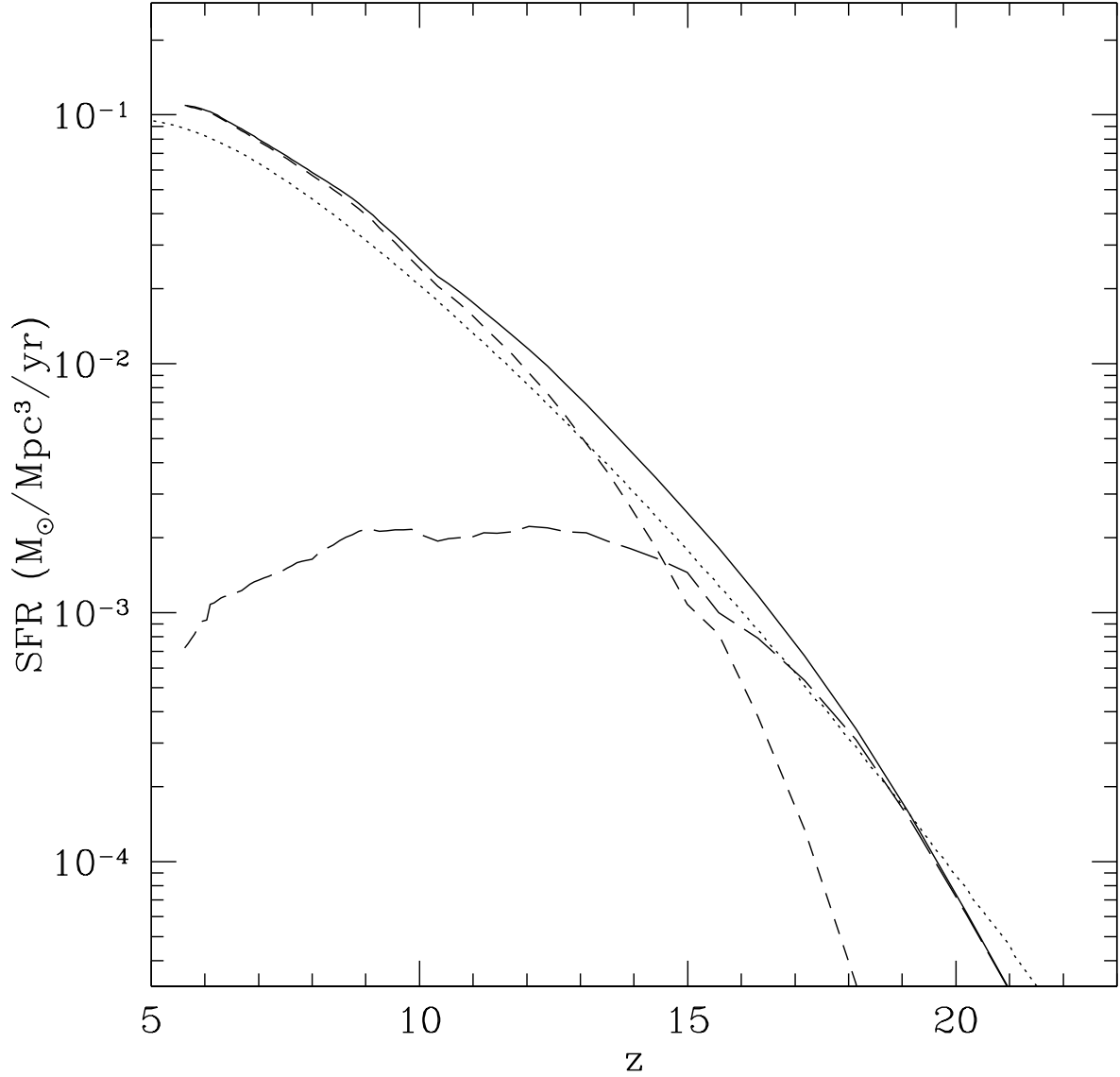}
\caption{Comoving star formation rate (SFR). The total SFR {\it (solid)}, from Pop III {\it (long dash)} 
and Pop II {\it (dash)} is consistent with that of \citet{hernquist03} {\it (dotted)}.}
\label{fig:star_formation_rate}
\end{figure}

\begin{figure}
\plotone{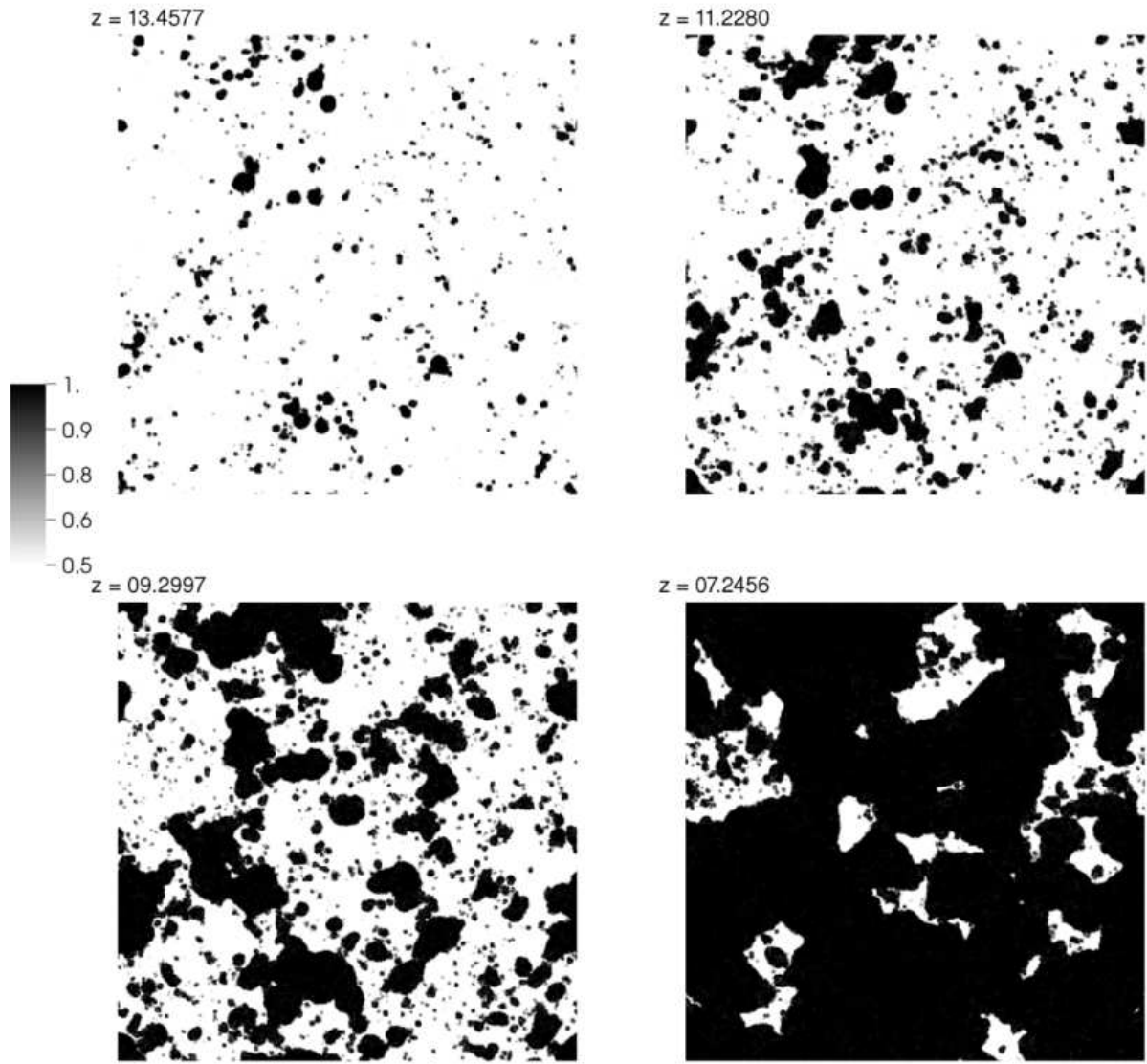}
\caption{
Distributions of ionization fraction at redshifts z\ $\sim$\ 13.5,\ 11.2,\ 9.3,\  
and 7.2. Highly ionized regions are represented by black while ionization fraction 
below 50 percent is shown as white. This is a plot of one slice in the simulation box. 
Each side of the plot is 100 Mpc/h. The global ionization fractions are $\sim$ 10, 30, 50, 
and 90 percents for z\ $\sim$\ 13.5,\ 11.2,\ 9.3,\ and 7.2, respectively.}
\label{fig:HII_evolution}
\end{figure}

\begin{figure}
\plottwo{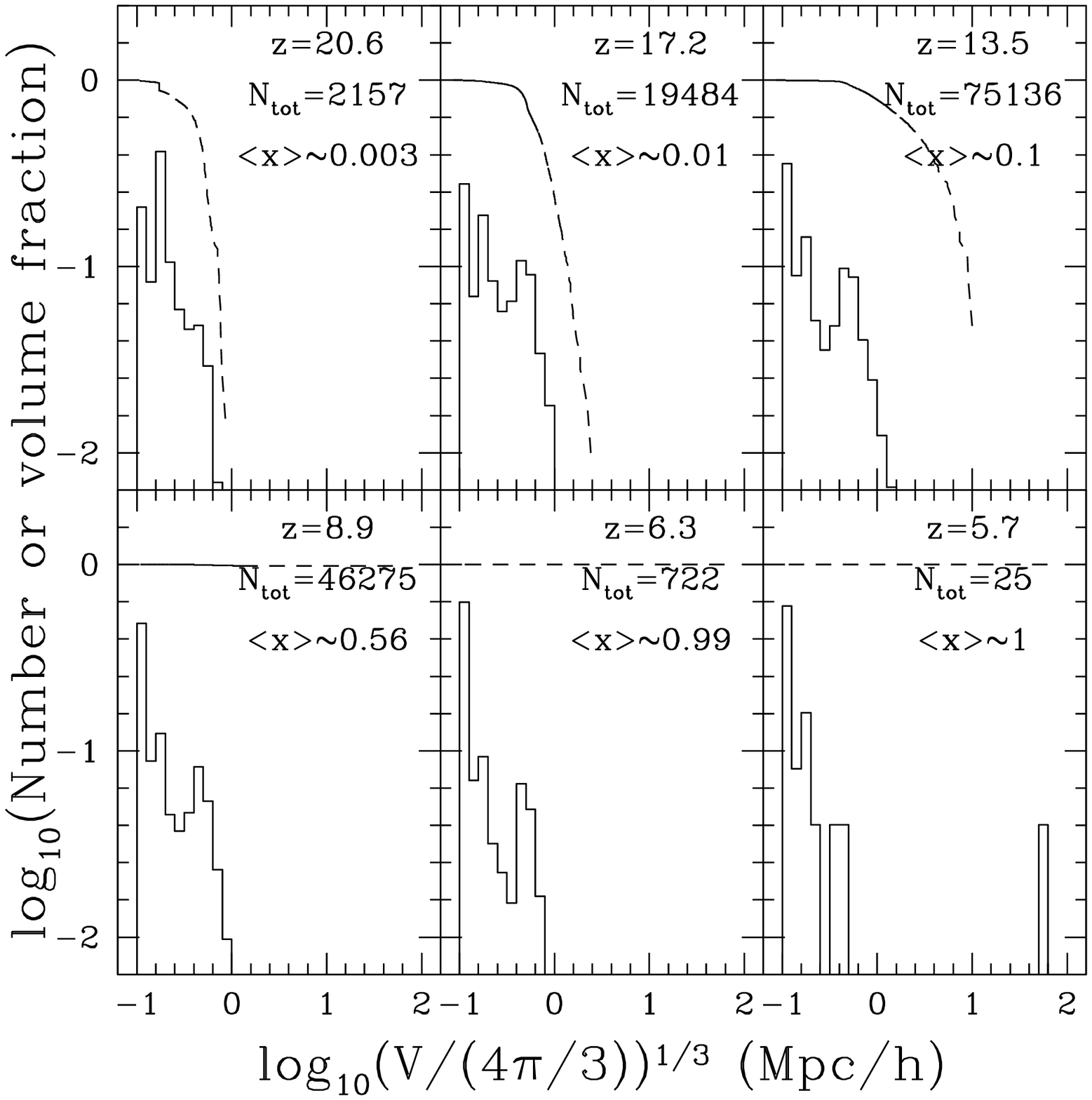}{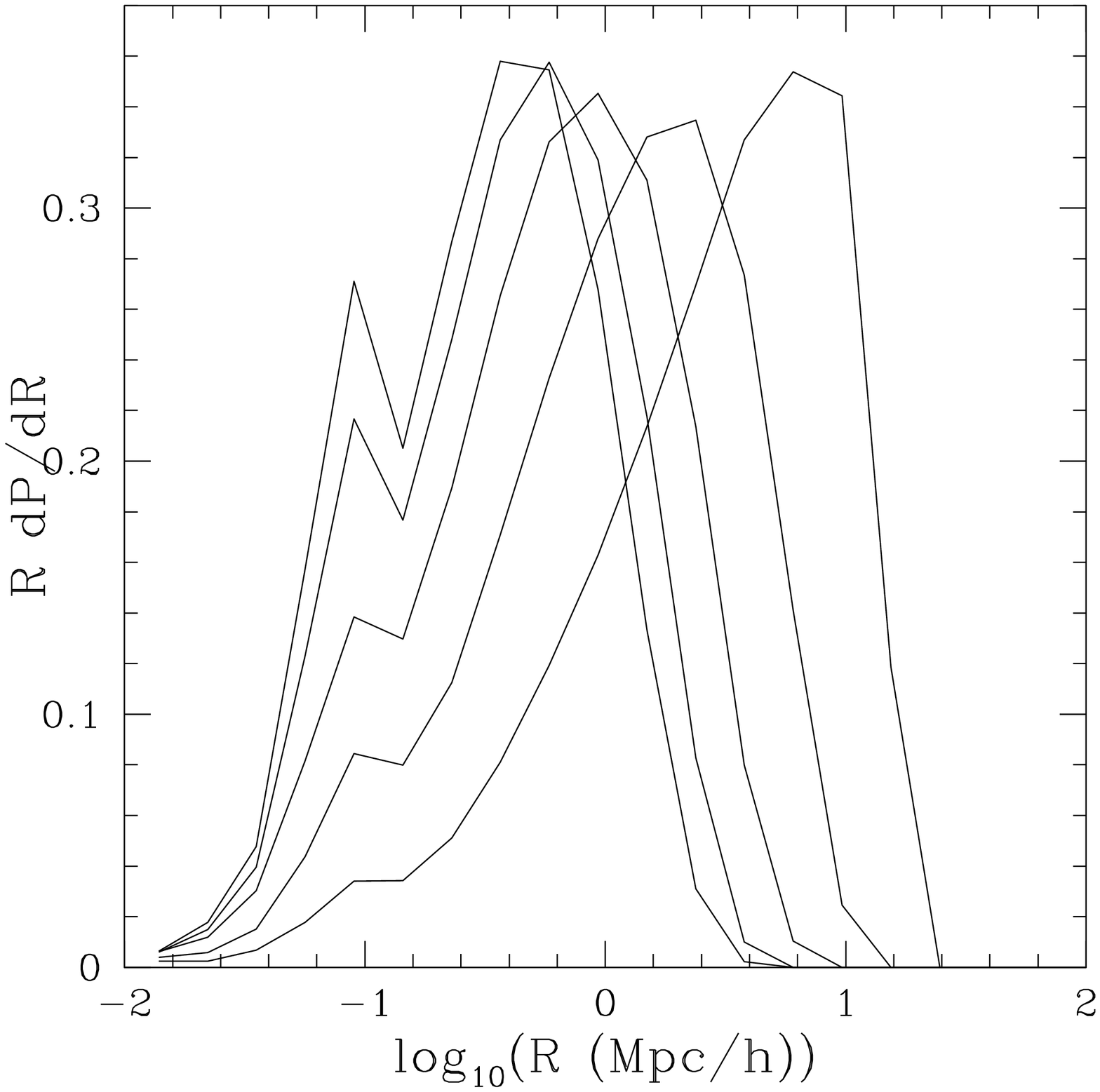}
\caption{Number and accumulated volume fraction distributions of HII 
bubble volumes {\it (left)} and bubble size distribution {\it (right)}. 
At redshifts z = 20.6, 17.2, 13.5, 8.9, 6.3, and 5.7, respectively, the number {\it (solid)} 
and accumulated volume distribution {\it (dash)} is given with 
the total number of bubbles, ${\rm N_{tot}}$, and the global volume-weighted mean ionization fraction, 
${\rm <x>}$. Before large bubbles that are comparable to the simulation box size appear around 
the redshift 6, bubble size distributions show three peaks that have the volumes 
of about 0.6, 0.03, and 0.006 ${\rm Mpc^{3}/h^{3}}$. The domination of one large bubble appears 
at z $<$ 10 as shown in the accumulated volume fractions of bubbles. In the right plot, 
we present the bubble size distribution that is measured as \citet{zahn07} at z = 
13.2, 12.1, 10.5, 9.0, and 8.0 that are correspondent to the global volume-weighted mean ionization 
fraction of 0.12, 0.18, 0.32, 0.52, and 0.75 from the small to large peak size.}
\label{fig:size_distribution}
\end{figure}

\begin{figure}
\plotone{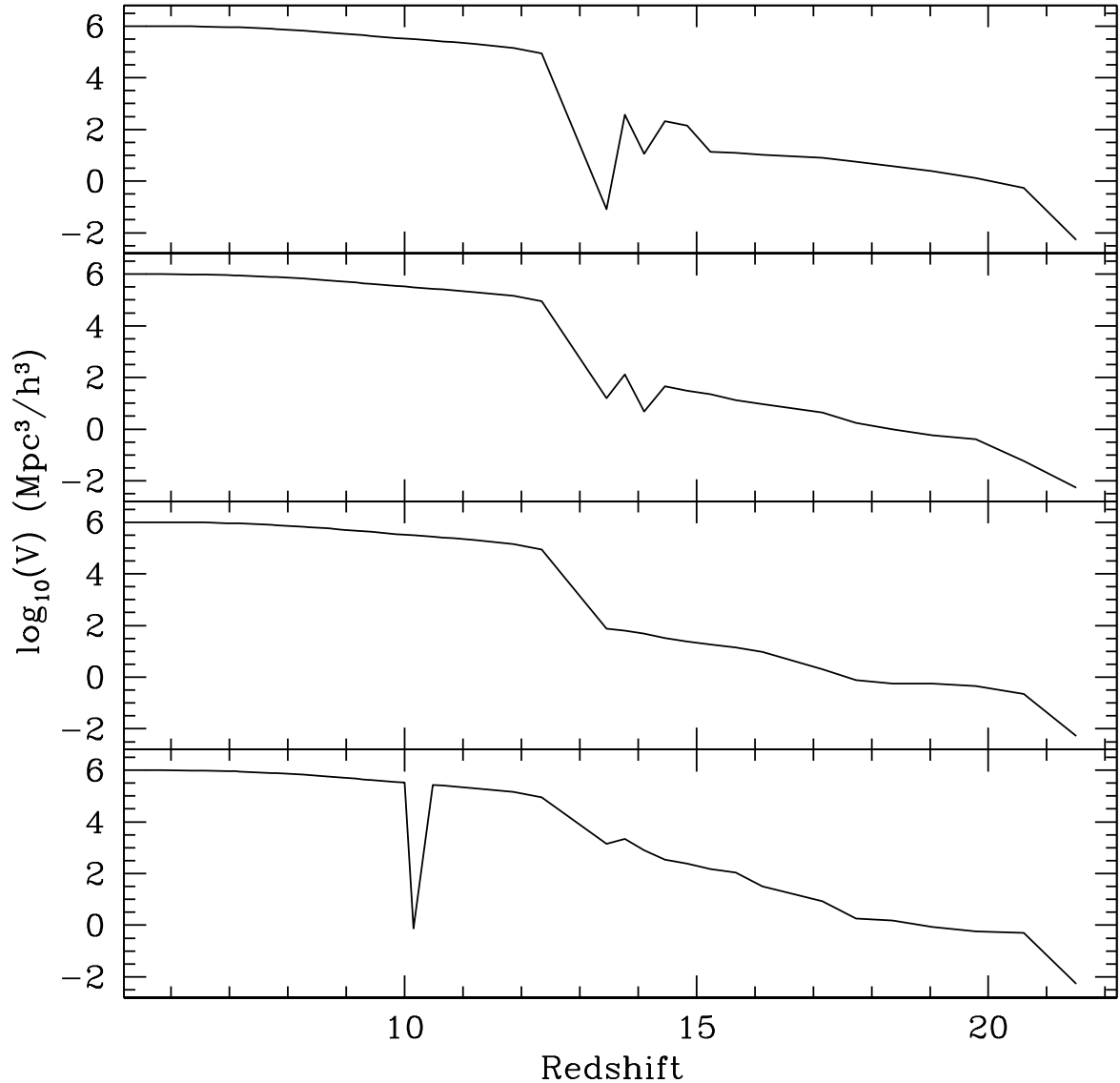}
\caption{
Bubble merger history. The evolution of the bubble size is given 
for four positions that are occupied by the four randomly selected 
smallest bubbles at z = 21.5. The bubbles that occupy the 
four positions are tracked and their volumes are measured. 
This bubble merger history is well explained by the change of 
a dominant stellar population and its ionizing photon production rate 
between z $\sim$ 10 and $\sim$ 15 as shown in Figure 7. During that 
transition bubbles experience reconnection to other bubbles.}
\label{fig:merger}
\end{figure}

\begin{figure}
\plotone{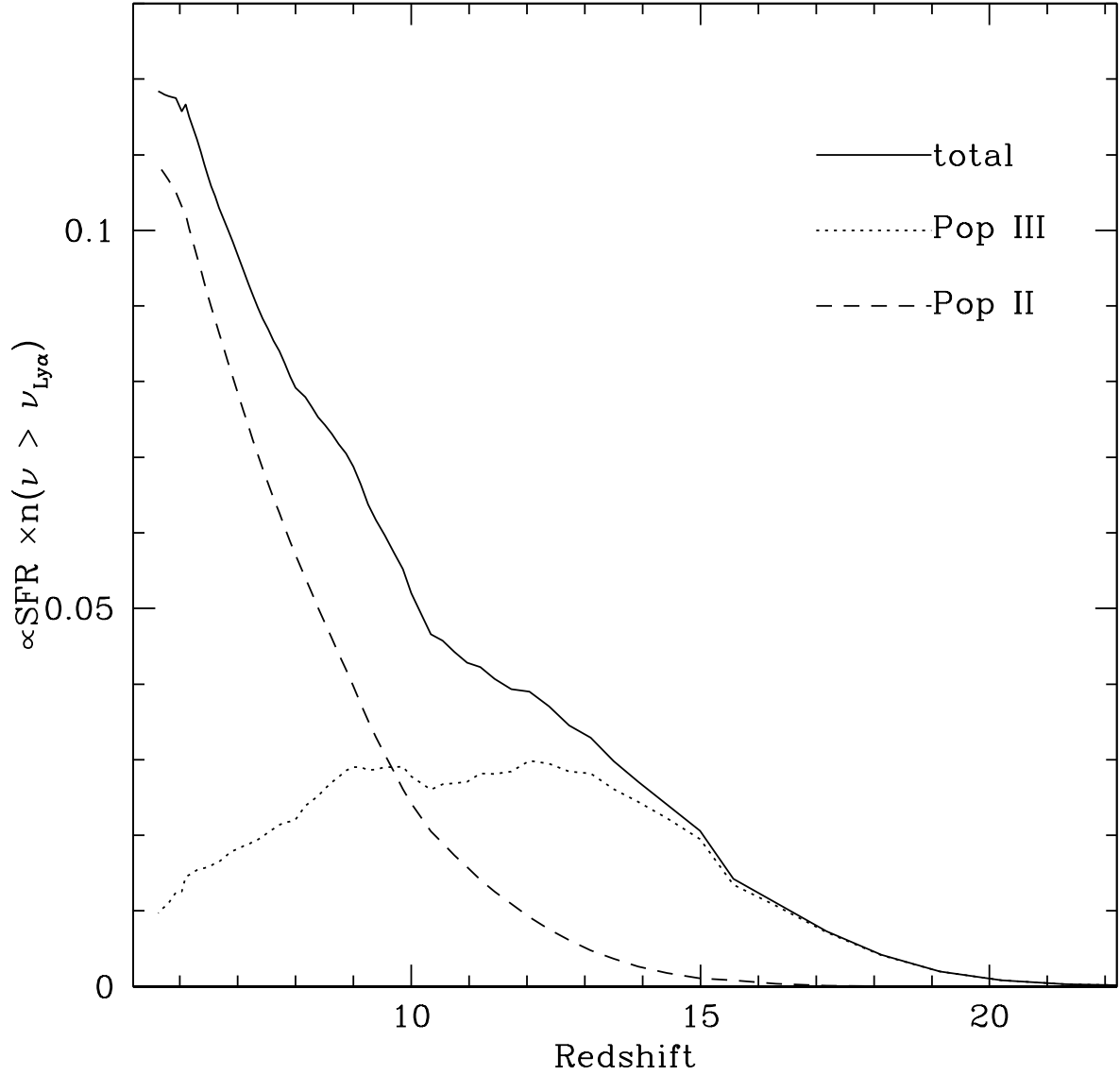}
\caption{Evolution of ionizing photon production rate from Pop III and Pop II stars. 
The ionizing photon production rate is estimated by a star forming rate times an 
ionizing photon of ${\rm h \nu \ge 13.6 eV}$ production rate per unit stellar mass.}
\label{fig:photon_production}
\end{figure}

\begin{figure}
\plotone{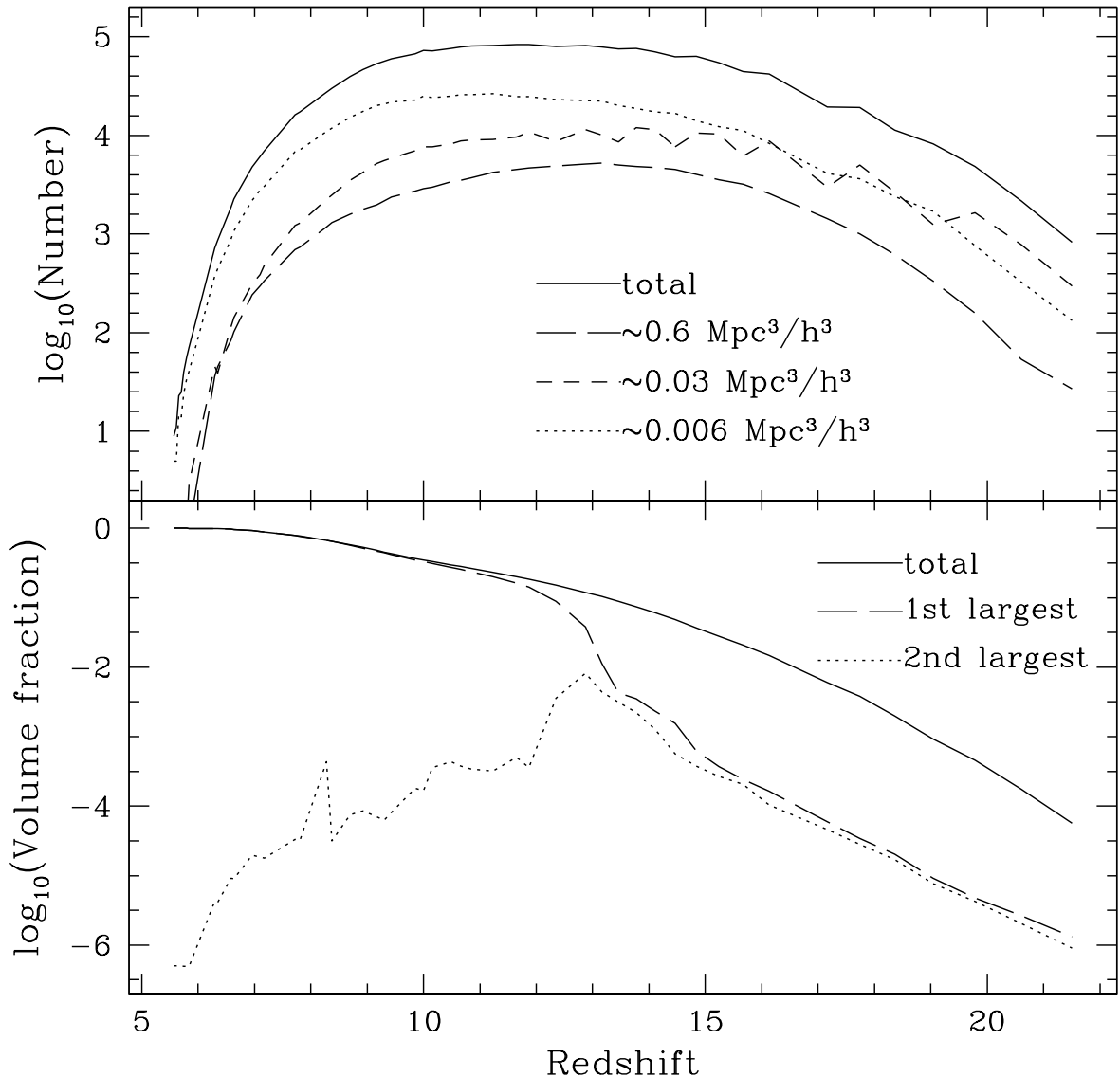}
\caption{Change of the number and volume fraction of HII bubbles. {\it(upper)} The total 
number of the HII bubbles increases until one large bubble dominates. Although the volumes of 
the smallest bubbles do not occupy a large fraction of the total volume, their number fraction 
dominates other sizes and shows a slightly later decline. {\it(lower)} Although the number 
of bubbles shows decrease after ${\rm z} \sim 10$, the total volume fraction of all HII bubbles 
increases as cosmic reionization continues. The volume fraction of the largest bubble 
at each epoch explains the change of the bubble number is mainly caused by the fact that 
intermediate-size bubbles are merged to the largest bubble so that even the second largest bubble 
occupy a small fraction of the total volume.}
\label{fig:number}
\end{figure}

\begin{figure}
\plotone{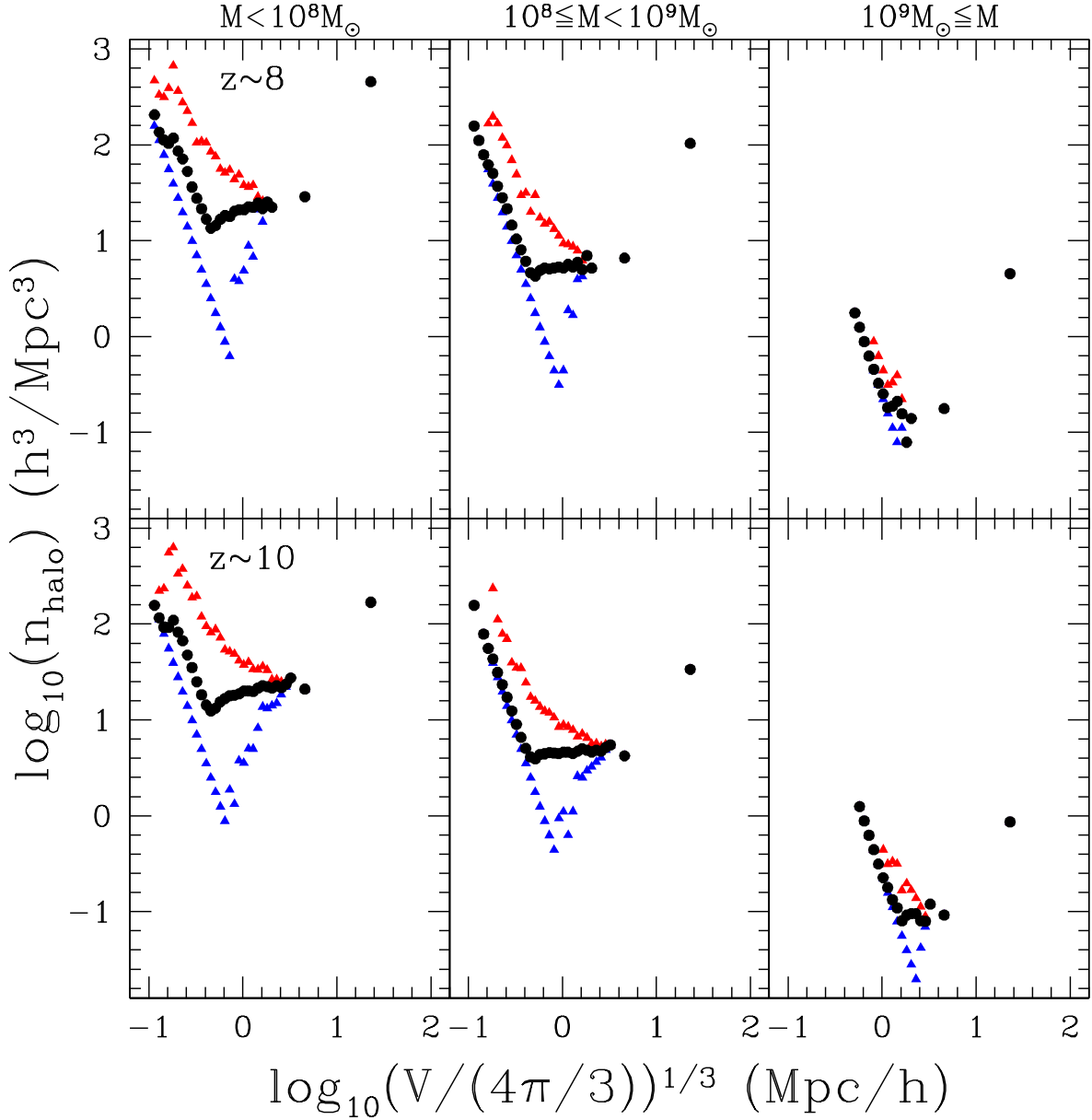}
\caption{
Number density distribution of dark matter halos inside HII bubbles at z $\sim$ 8 and 10. Open triangle 
(red), 
dot (black), and open rectangle (blue) represent maximum, average, and minimum number density of halos for 
a given bubble size, respectively. The ranges of DM halo masses 
are ${\rm M_{halo} < 10^{8} ~ M_{\odot}/h, 10^{8}\ M_{\odot}/h \leq M_{halo} < 10^{9}\ M_{\odot}/h}$, 
and ${\rm 10^{9}\ M_{\odot}/h \leq M_{halo}}$. [{\it See the electronic edition of the Journal 
for a color version of this figure.}]}
\label{fig:halo_size}
\end{figure}

\begin{figure}
\plotone{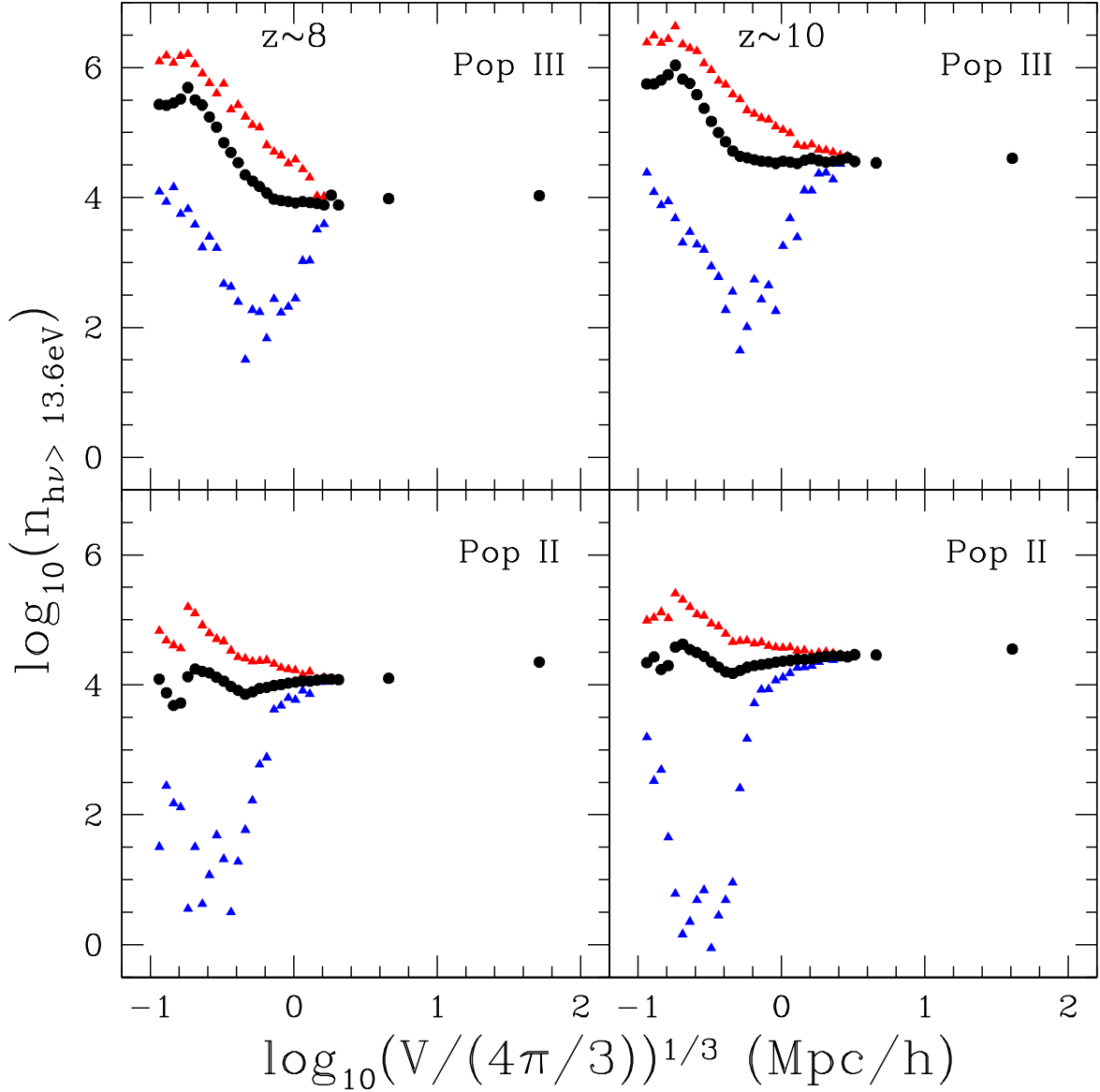}
\caption{
Ionizing photon production rate per unit volume inside HII bubbles at redshifts 8.3 and 10.2. 
The ionizing photon production rate is calculated as used in Figure 6. 
Open triangle (red), dot (black), and open rectangle (blue) data points 
are maximum, average, and minimum values for different sizes of bubbles, respectively. Here we 
plot only bubbles that have stellar masses inside them. Note that ionizing photon production 
rate of Pop II stars is comparable to that of Pop III stars for bubbles of 
${\rm V\ \simgt\ 0.6\ Mpc^{3}/h^{3}}$. 
[{\it See the electronic edition of the Journal for a color version of this figure.}]}
\label{fig:photon_size}
\end{figure}

\begin{figure}
\plotone{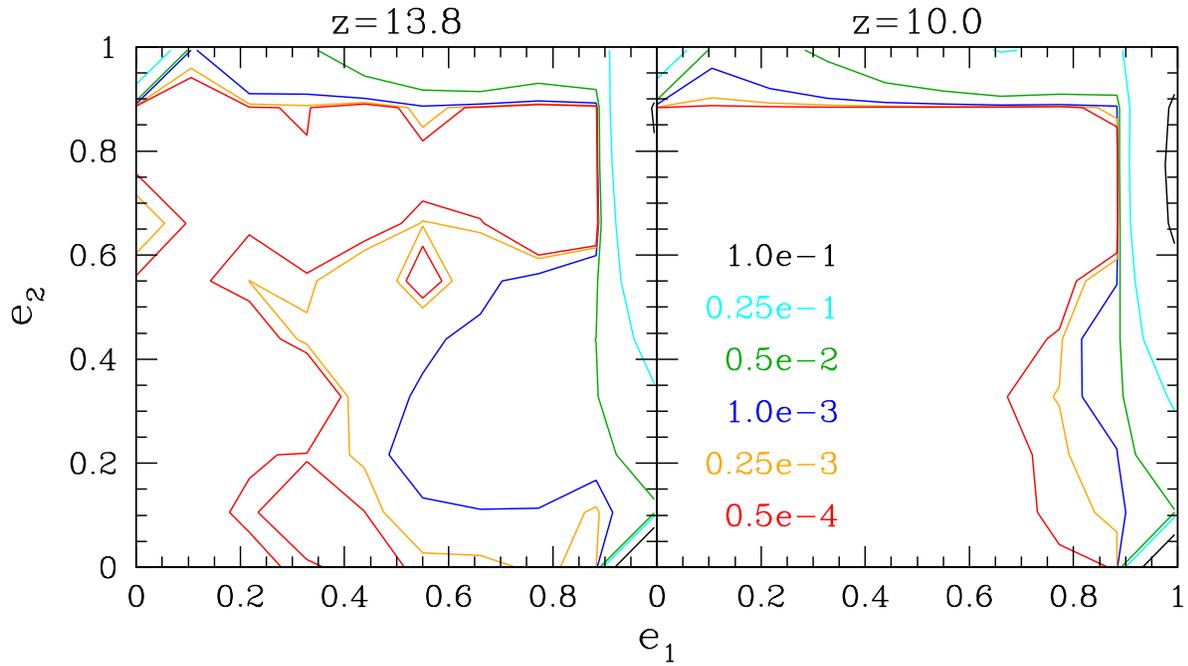}
\caption{Distribution probability of the HII bubble shapes expressed as $e_{1}$ and $e_{2}$ at z = 13.8 and 6.3. 
Most bubbles are far from a sphere, i.e. $e_{1}$ and $e_{2} \gg 0$. This trend does not change even though 
reionized volume becomes increasing. The distribution is derived from Equation 5. 
[{\it See the electronic edition of the Journal for a color version of this figure.}]}
\label{fig:bubble_shape}
\end{figure}

\begin{figure}
\plotone{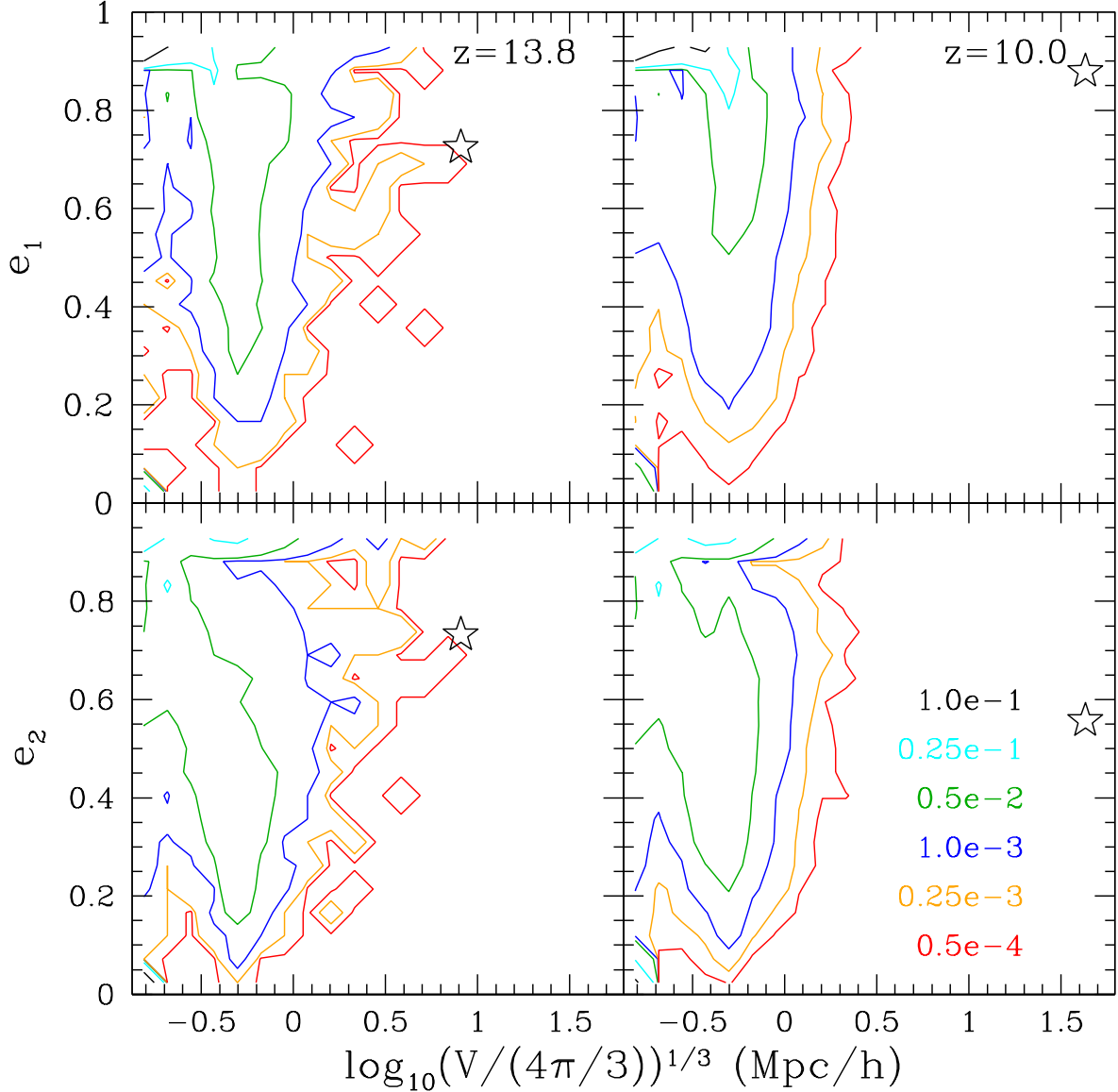}
\caption{
Distribution probability of bubble shape parameters and volume at z = 13.8 and 10. The bubble shape 
has a distribution that does not change even though the fraction of 
reionized regions increases between z = 13.8 and 10. While small bubbles have the narrow range of shapes 
around $e_{1}$ or $e_{2} \sim 0$ or 1, the bubbles of intermediate size show 
the broad range of shapes. The star symbols represent the shape of the largest bubble. 
[{\it See the electronic edition of the Journal for a color version of this figure.}]}
\label{fig:shape_size}
\end{figure}

\begin{figure}
\plotone{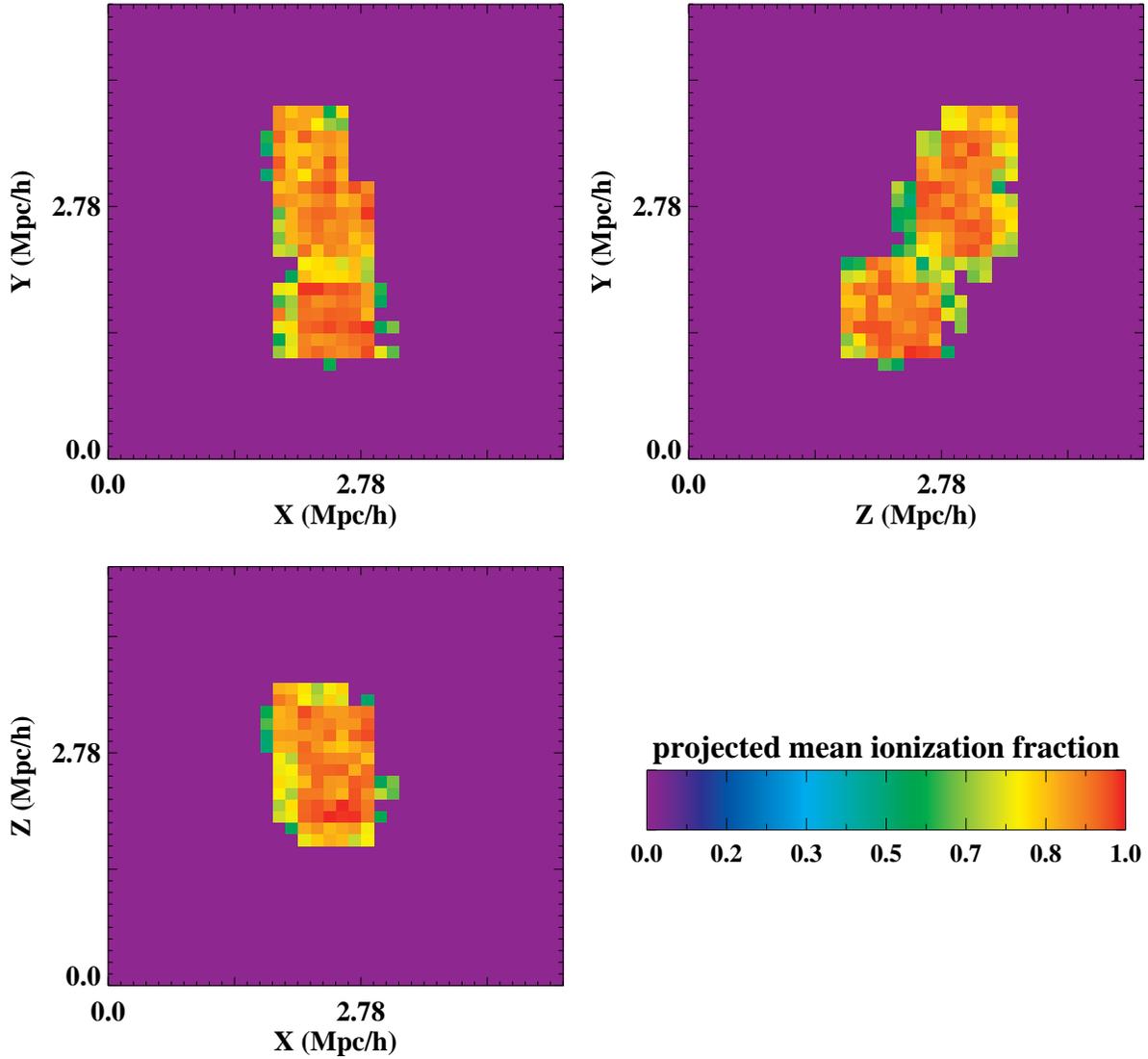}
\caption{Projected mean ionization fraction distribution of a single bubble at z = 13.8. 
Its derived shape parameters, $e_{1} \sim 0.93$ and $e_{2} \sim 0.5$, are consistent with 
this ionization fraction distribution. Here we set the background around the bubble to have 
zero ionization fraction for visualization.}
\label{fig:one_bubble}
\end{figure}

\begin{figure}
\plotone{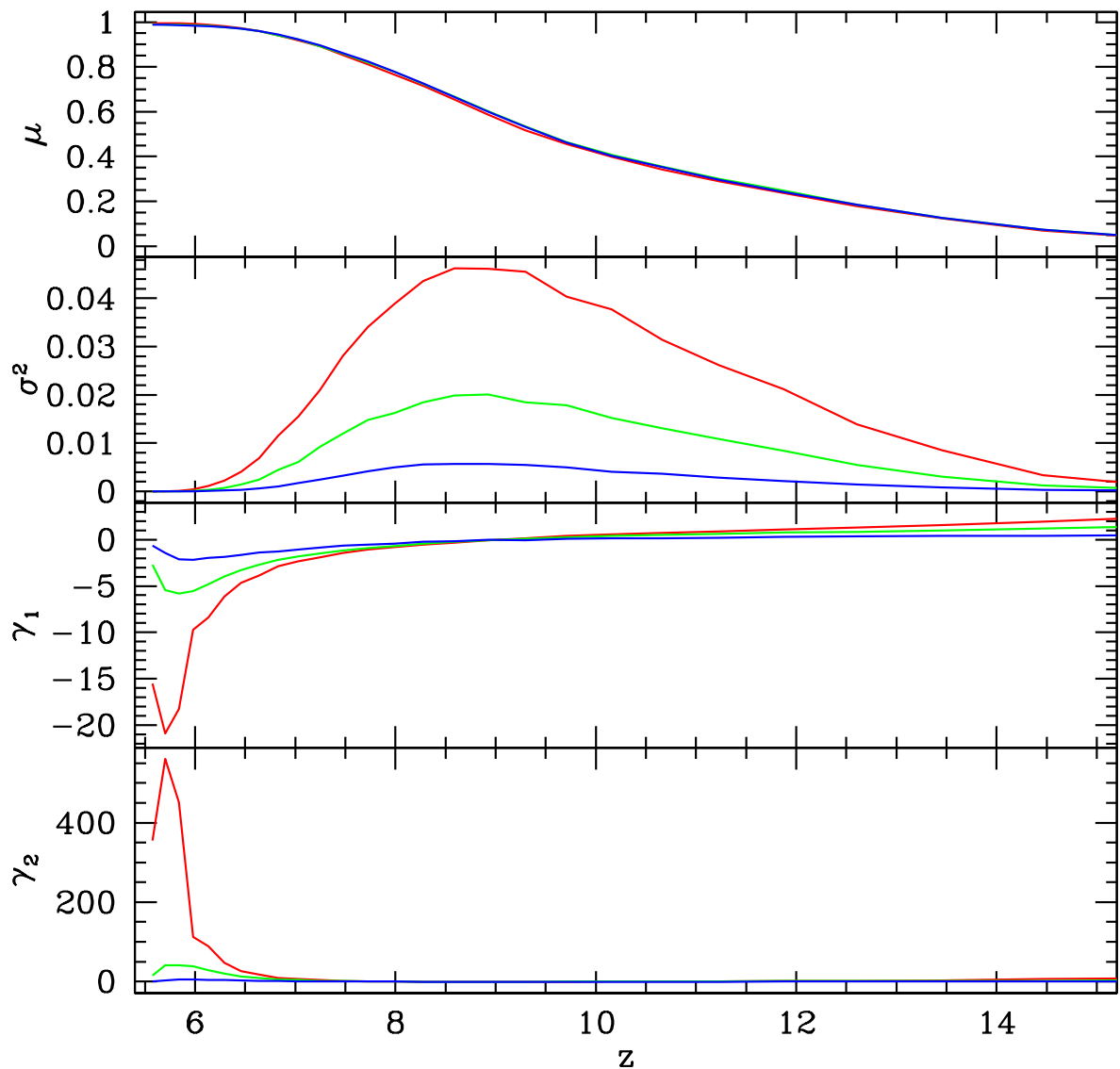}
\caption{Non-Gaussianity of reionization field. The statistics of mass-weighted ionization fraction 
are given for mean, variance, skewness, and kurtosis from top to bottom. Each statistic is calculated 
by sampling 5000 spheres of the comoving radius $5$ {\it (dot line, red)}, $10$ 
{\it (dash line, green)}, and $20$ {\it (solid line, blue)} 
Mpc/h, respectively. 
[{\it See the electronic edition of the Journal for a color version of this figure.}]}
\label{fig:non_gaussianity}
\end{figure}

\begin{figure}
\plotone{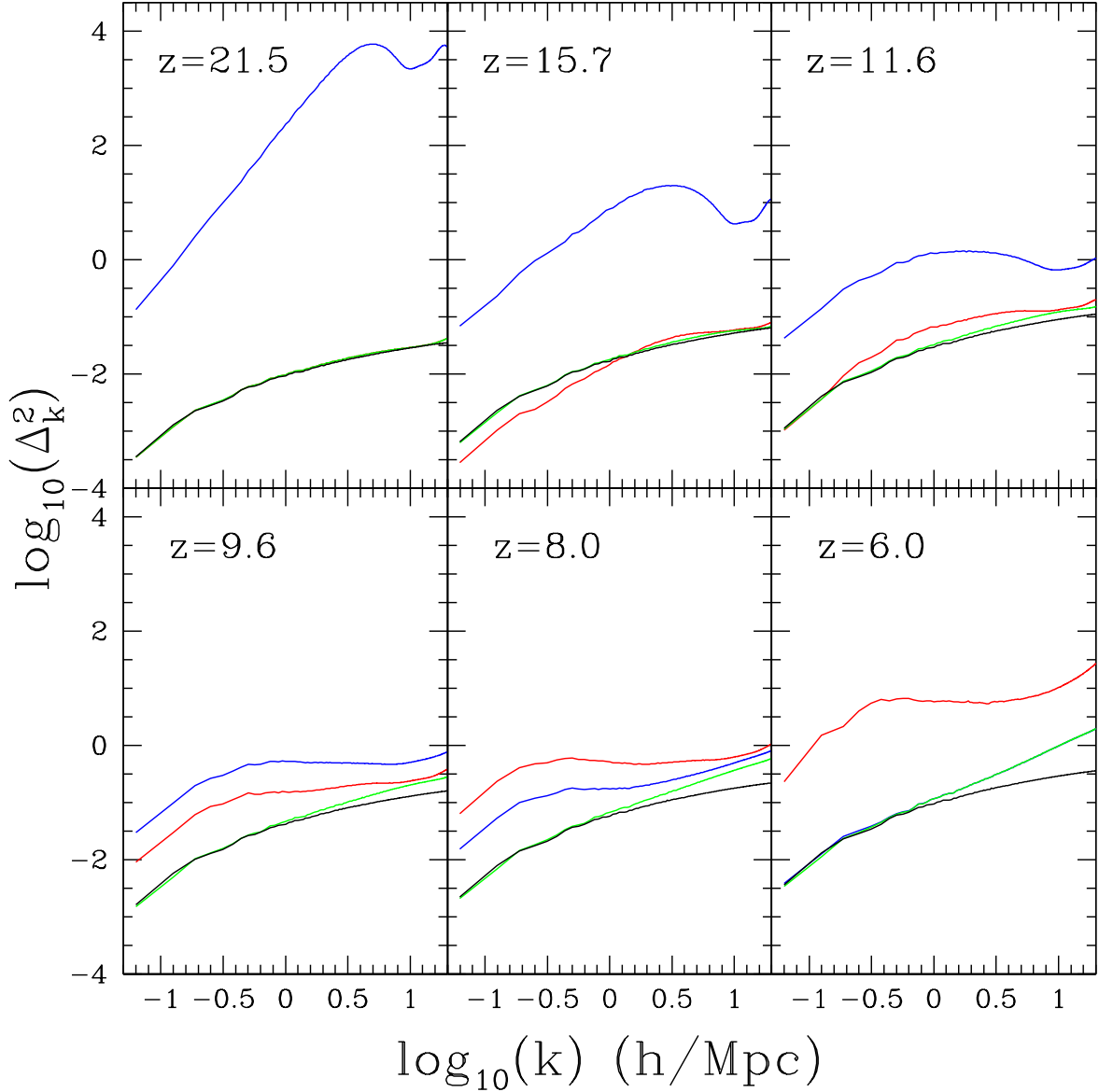}
\caption{Power spectrum of ${\rm \delta_{HI}}$, ${\rm \delta_{HII}}$, and 
${\rm \delta_{mat}}$. At high redshift the power spectrum of a neutral hydrogen 
{\it(dot line, red)} is well matched to that of matter {\it(dash line, green)} 
while an ionized hydrogen 
{\it(dot-dash line, blue)} shows the match to the matter power spectrum at low redshift. A 
linearly developed matter power spectrum {\it(solid line, black)} does not match to a 
matter power spectrum at a small scale as a non-linear structure forms. 
[{\it See the electronic edition of the Journal for a color version of this figure.}]}
\label{fig:power_spectrum}
\end{figure}

\begin{figure}
\plotone{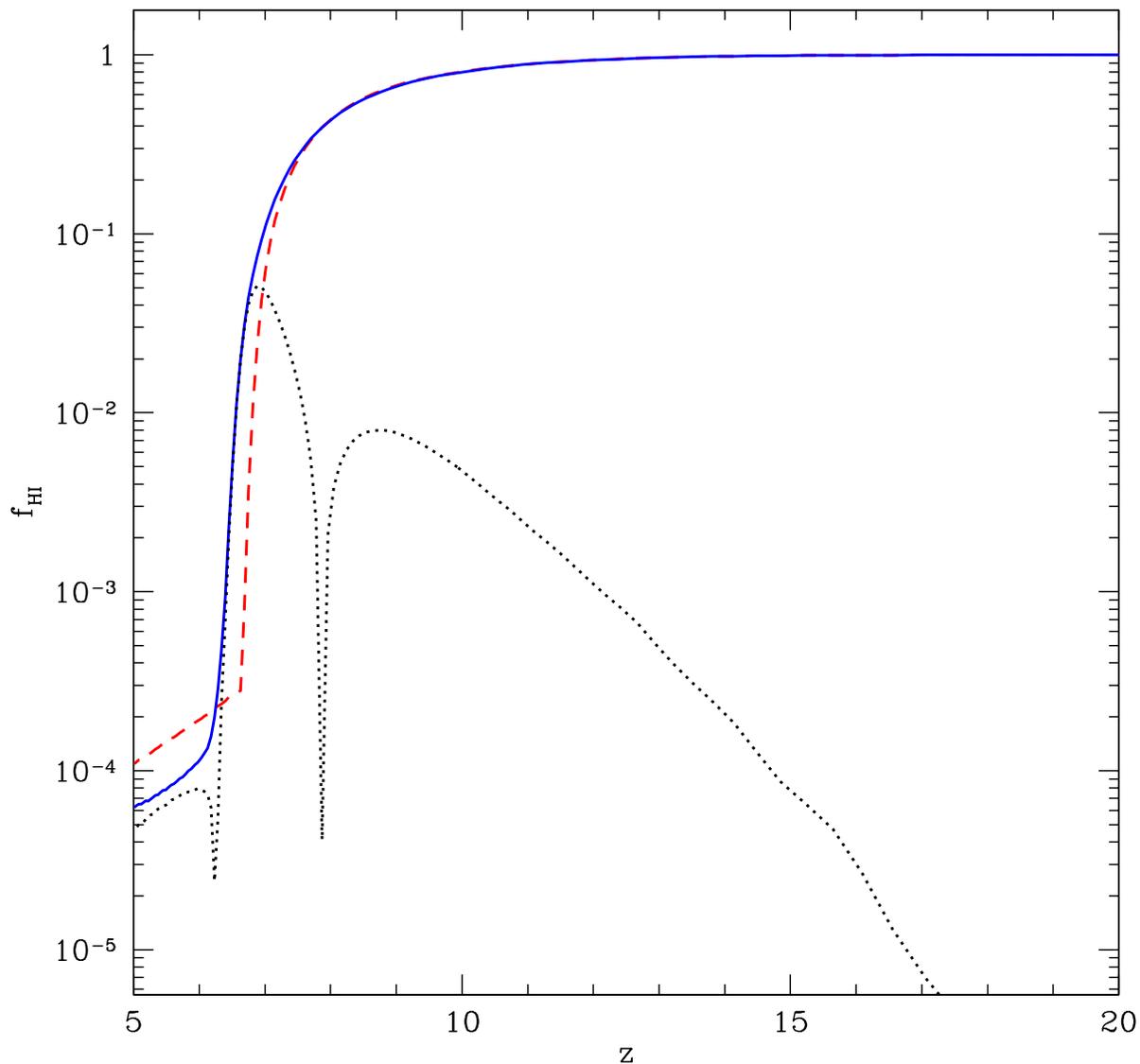}
\caption{A comparison of the volume-weighted neutral hydrogen fraction $f_{\rm HI}$ 
from the photon-advection scheme (blue, solid) and ray-tracing scheme (red, dashed) 
for radiative transfer. The photon-advection scheme correctly captures the reionization 
process up until it is $\sim75\%$ completed by volume. The magnitude of the difference 
in $f_{\rm HI}$ (black, dotted) is generally very small and only reaches a maximum value 
of 0.05 near complete overlap. [{\it See the electronic edition of the Journal 
for a color version of this figure.}]}
\end{figure}

\begin{figure}
\plotone{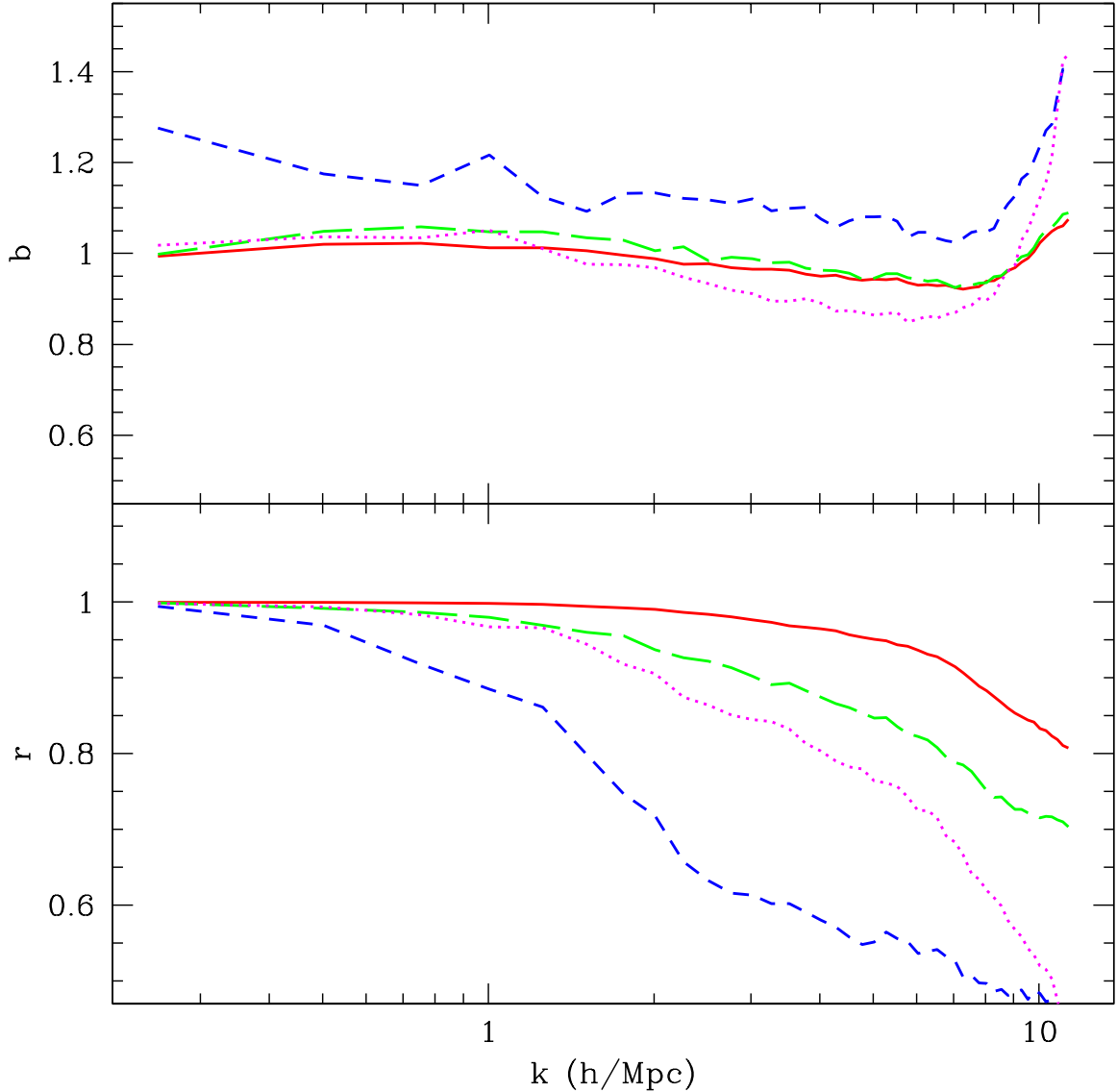}
\caption{A comparison of the photon-advection scheme and ray-tracing scheme for radiative 
transfer using the bias $b(k)$ and cross-correlation $r(k)$ of the neutral hydrogen 
fraction $f_{\rm HI}(x)$ field.  At $z=8.2$ (red, solid) and 7.5 (green, long-dashed), 
both simulations have volume-averaged neutral fractions of 0.5 and 0.25, respectively, and 
the spatial evolution of HI is very similar. At $=7.1$ (blue, short-dashed), the correlation 
is poor because reionization has progressed slightly faster in the ray-tracing scheme. 
However, when the two simulations are compared at the same neutral fraction of 0.1 
(magenta, dotted), the correlation is much better, particularly on large scales. 
[{\it See the electronic edition of the Journal 
for a color version of this figure.}]}
\end{figure}

\end{document}